\documentclass[%
 reprint,
superscriptaddress,
preprintnumbers,
nofootinbib,
 amsmath,amssymb,
 aps,
 prd,
]{revtex4-2}
\usepackage{graphicx}
\usepackage{dcolumn}
\usepackage{bm}
\usepackage{hyperref}
\usepackage{mathtools}

\usepackage{booktabs}
\usepackage{cleveref}

\usepackage{xcolor}

\usepackage{subfiles}

\usepackage{siunitx}
\newcolumntype{d}[1]{D{.}{.}{#1}}

\usepackage{lineno}

\defcitealias{Bocquet24Ia}{SB24a}

\defcitealias{Bocquet24II}{SB24b}

\defcitealias{Vogt24}{SV24a}

\usepackage{hyperref}
\definecolor{plasmablue}{rgb}{0.050383, 0.029803, 0.527975}
\hypersetup{
  colorlinks=true,
  urlcolor=plasmablue,
  citecolor=plasmablue,
  }

\newcommand{\ie}{i.\,e.\ }
\newcommand{\eg}{e.\,g.\ }

\newcommand{\dd}{\textnormal{d}}

\newcommand{\logfR}{\ensuremath{{\log_{10}|f_{R0}|}}}

\newcommand{\zetahat}{\ensuremath{{\hat \zeta}}}
\newcommand{\lambdahat}{\ensuremath{{\hat \lambda}}}

\newcommand{\asz}{\ensuremath{{\zeta_0}}}
\newcommand{\bsz}{\ensuremath{{\zeta_M}}}
\newcommand{\csz}{\ensuremath{{\zeta_z}}}
\newcommand{\sigmalnzeta}{\ensuremath{{\sigma_{\ln\zeta}}}}
\newcommand{\alambda}{\ensuremath{{\lambda_0}}}
\newcommand{\blambda}{\ensuremath{{\lambda_M}}}
\newcommand{\clambda}{\ensuremath{{\lambda_z}}}
\newcommand{\sigmalnlambda}{\ensuremath{{\sigma_{\ln\lambda}}}}
\newcommand{\MWL}{\ensuremath{M_\mathrm{WL}}}
\newcommand{\bWL}{\ensuremath{{\ln M_{\mathrm{WL}_0}}}}
\newcommand{\bWLM}{\ensuremath{{M_{\mathrm{WL}_M}}}}
\newcommand{\sWLall}{\ensuremath{{\sigma_{\ln\mathrm{WL}}}}}
\newcommand{\sWL}{\ensuremath{{\ln\sigma^2_{\ln\mathrm{WL}_0}}}}
\newcommand{\sWLM}{\ensuremath{\sigma^2_{\ln\mathrm{WL}_M}}}

\newcommand{\bWLHST}{\ensuremath{{\ln M_{\mathrm{WL}_0}}}}
\newcommand{\sWLHST}{\ensuremath{{\sigma_{\ln\mathrm{WL}}}}}

\begin{document}

\preprint{DES-2024-0856}
\preprint{FERMILAB-PUB-24-0580-PPD}

\title{Constraints on \texorpdfstring{$f(R)$}{f(R)} gravity from tSZE-selected SPT galaxy clusters and weak lensing mass calibration from DES and HST}

\author{S.~M.~L.~Vogt}
\email{s.vogt@physik.lmu.de}
\affiliation{
University Observatory, Faculty of Physics, Ludwig-Maximilians-Universit\"at, Scheinerstr.\ 1, 81679 Munich, Germany
}
\affiliation{
Excellence Cluster Origins, Boltzmannstr. 2, 85748 Garching, Germany
}
\affiliation{Max Planck Institute for Astrophysics, Karl-Schwarzschild-Str. 1, 85748 Garching, Germany}
\author{S.~Bocquet}
\affiliation{
University Observatory, Faculty of Physics, Ludwig-Maximilians-Universit\"at, Scheinerstr.\ 1, 81679 Munich, Germany
}
\author{C.~T.~Davies}
\affiliation{
University Observatory, Faculty of Physics, Ludwig-Maximilians-Universit\"at, Scheinerstr.\ 1, 81679 Munich, Germany
}
\author{J.~J.~Mohr}
\affiliation{
University Observatory, Faculty of Physics, Ludwig-Maximilians-Universit\"at, Scheinerstr.\ 1, 81679 Munich, Germany
}
\affiliation{Max Planck Institute for Extraterrestrial Physics, Giessenbachstr. 2, 85748 Garching, Germany}
\author{F.~Schmidt}
\affiliation{Max Planck Institute for Astrophysics, Karl-Schwarzschild-Str. 1, 85748 Garching, Germany}

\author{C.-Z.~Ruan}
\affiliation{Institute of Theoretical Astrophysics, University of Oslo, 0315 Oslo, Norway}

\author{B.~Li}
\affiliation{Institute for Computational Cosmology, Department of Physics, Durham University, South Road, Durham DH1 3LE, UK}

\author{C.~Hernández-Aguayo}
\affiliation{Max Planck Institute for Astrophysics, Karl-Schwarzschild-Str. 1, 85748 Garching, Germany}

\author{S.~Grandis}
\affiliation{Universit\"at Innsbruck, Institut f\"ur Astro- und Teilchenphysik, Technikerstr. 25/8, 6020 Innsbruck, Austria}
\affiliation{
University Observatory, Faculty of Physics, Ludwig-Maximilians-Universit\"at, Scheinerstr.\ 1, 81679 Munich, Germany
}
\author{L.~E.~Bleem}
\affiliation{High-Energy Physics Division, Argonne National Laboratory, 9700 South Cass Avenue, Lemont, IL 60439, USA}
\affiliation{Kavli Institute for Cosmological Physics, University of Chicago, 5640 South Ellis Avenue, Chicago, IL 60637, USA}
\author{M.~Klein}
\affiliation{
University Observatory, Faculty of Physics, Ludwig-Maximilians-Universit\"at, Scheinerstr.\ 1, 81679 Munich, Germany
}
\affiliation{Max Planck Institute for Extraterrestrial Physics, Giessenbachstr. 2, 85748 Garching, Germany}
\author{T.~Schrabback}
\affiliation{Universit\"at Innsbruck, Institut f\"ur Astro- und Teilchenphysik, Technikerstr. 25/8, 6020 Innsbruck, Austria}

\author{M.~Aguena}
\affiliation{Laborat\'orio Interinstitucional de e-Astronomia - LIneA, Rua Gal. Jos\'e Cristino 77, Rio de Janeiro, RJ - 20921-400, Brazil}
\author{D.~Brooks}
\affiliation{Department of Physics \& Astronomy, University College London, Gower Street, London, WC1E 6BT, UK}
\author{D.~L.~Burke}
\affiliation{Kavli Institute for Particle Astrophysics \& Cosmology, P. O. Box 2450, Stanford University, Stanford, CA 94305, USA}
\affiliation{SLAC National Accelerator Laboratory, Menlo Park, CA 94025, USA}
\author{A.~Campos}
\affiliation{Department of Physics, Carnegie Mellon University, Pittsburgh, Pennsylvania 15312, USA}
\affiliation{NSF AI Planning Institute for Physics of the Future, Carnegie Mellon University, Pittsburgh, PA 15213, USA}
\author{A.~Carnero~Rosell}
\affiliation{Instituto de Astrofisica de Canarias, E-38205 La Laguna, Tenerife, Spain}
\affiliation{Laborat\'orio Interinstitucional de e-Astronomia - LIneA, Rua Gal. Jos\'e Cristino 77, Rio de Janeiro, RJ - 20921-400, Brazil}
\author{J.~Carretero}
\affiliation{Institut de F\'{\i}sica d'Altes Energies (IFAE), The Barcelona Institute of Science and Technology, Campus UAB, 08193 Bellaterra (Barcelona) Spain}
\author{M.~Costanzi}
\affiliation{Astronomy Unit, Department of Physics, University of Trieste, via Tiepolo 11, I-34131 Trieste, Italy}
\affiliation{INAF-Osservatorio Astronomico di Trieste, via G. B. Tiepolo 11, I-34143 Trieste, Italy}
\affiliation{Institute for Fundamental Physics of the Universe, Via Beirut 2, 34014 Trieste, Italy}
\author{L.~N.~da Costa}
\affiliation{Laborat\'orio Interinstitucional de e-Astronomia - LIneA, Rua Gal. Jos\'e Cristino 77, Rio de Janeiro, RJ - 20921-400, Brazil}
\author{M.~E.~S.~Pereira}
\affiliation{Hamburger Sternwarte, Universit\"{a}t Hamburg, Gojenbergsweg 112, 21029 Hamburg, Germany}
\author{J.~De~Vicente}
\affiliation{Centro de Investigaciones Energ\'eticas, Medioambientales y Tecnol\'ogicas (CIEMAT), Madrid, Spain}
\author{P.~Doel}
\affiliation{Department of Physics \& Astronomy, University College London, Gower Street, London, WC1E 6BT, UK}
\author{S.~Everett}
\affiliation{California Institute of Technology, 1200 East California Blvd, MC 249-17, Pasadena, CA 91125, USA}
\author{I.~Ferrero}
\affiliation{Institute of Theoretical Astrophysics, University of Oslo. P.O. Box 1029 Blindern, NO-0315 Oslo, Norway}
\author{J.~Frieman}
\affiliation{Fermi National Accelerator Laboratory, P. O. Box 500, Batavia, IL 60510, USA}
\affiliation{Kavli Institute for Cosmological Physics, University of Chicago, Chicago, IL 60637, USA}
\author{J.~Garc\'ia-Bellido}
\affiliation{Instituto de Fisica Teorica UAM/CSIC, Universidad Autonoma de Madrid, 28049 Madrid, Spain}
\author{M.~Gatti}
\affiliation{Department of Physics and Astronomy, University of Pennsylvania, Philadelphia, PA 19104, USA}
\author{G.~Giannini}
\affiliation{Institut de F\'{\i}sica d'Altes Energies (IFAE), The Barcelona Institute of Science and Technology, Campus UAB, 08193 Bellaterra (Barcelona) Spain}
\affiliation{Kavli Institute for Cosmological Physics, University of Chicago, Chicago, IL 60637, USA}
\author{D.~Gruen}
\affiliation{
University Observatory, Faculty of Physics, Ludwig-Maximilians-Universit\"at, Scheinerstr.\ 1, 81679 Munich, Germany
}
\author{R.~A.~Gruendl}
\affiliation{Center for Astrophysical Surveys, National Center for Supercomputing Applications, 1205 West Clark St., Urbana, IL 61801, USA}
\affiliation{Department of Astronomy, University of Illinois at Urbana-Champaign, 1002 W. Green Street, Urbana, IL 61801, USA}
\author{S.~R.~Hinton}
\affiliation{School of Mathematics and Physics, University of Queensland,  Brisbane, QLD 4072, Australia}
\author{D.~L.~Hollowood}
\affiliation{Santa Cruz Institute for Particle Physics, Santa Cruz, CA 95064, USA}
\author{S.~Lee}
\affiliation{Jet Propulsion Laboratory, California Institute of Technology, 4800 Oak Grove Dr., Pasadena, CA 91109, USA}
\author{M.~Lima}
\affiliation{Departamento de F\'isica Matem\'atica, Instituto de F\'isica, Universidade de S\~ao Paulo, CP 66318, S\~ao Paulo, SP, 05314-970, Brazil}
\affiliation{Laborat\'orio Interinstitucional de e-Astronomia - LIneA, Rua Gal. Jos\'e Cristino 77, Rio de Janeiro, RJ - 20921-400, Brazil}
\author{J.~L.~Marshall}
\affiliation{George P. and Cynthia Woods Mitchell Institute for Fundamental Physics and Astronomy, and Department of Physics and Astronomy, Texas A\&M University, College Station, TX 77843,  USA}
\author{J. Mena-Fern{\'a}ndez}
\affiliation{LPSC Grenoble - 53, Avenue des Martyrs 38026 Grenoble, France}
\author{R.~Miquel}
\affiliation{Instituci\'o Catalana de Recerca i Estudis Avan\c{c}ats, E-08010 Barcelona, Spain}
\affiliation{Institut de F\'{\i}sica d'Altes Energies (IFAE), The Barcelona Institute of Science and Technology, Campus UAB, 08193 Bellaterra (Barcelona) Spain}
\author{J.~Myles}
\affiliation{Department of Astrophysical Sciences, Princeton University, Peyton Hall, Princeton, NJ 08544, USA}
\author{M.~Paterno}
\affiliation{Fermi National Accelerator Laboratory, P. O. Box 500, Batavia, IL 60510, USA}
\author{A.~Pieres}
\affiliation{Laborat\'orio Interinstitucional de e-Astronomia - LIneA, Rua Gal. Jos\'e Cristino 77, Rio de Janeiro, RJ - 20921-400, Brazil}
\affiliation{Observat\'orio Nacional, Rua Gal. Jos\'e Cristino 77, Rio de Janeiro, RJ - 20921-400, Brazil}
\author{A.~A.~Plazas~Malag\'on}
\affiliation{Kavli Institute for Particle Astrophysics \& Cosmology, P. O. Box 2450, Stanford University, Stanford, CA 94305, USA}
\affiliation{SLAC National Accelerator Laboratory, Menlo Park, CA 94025, USA}
\author{C.~L.~Reichardt}
\affiliation{School of Physics, University of Melbourne, Parkville, VIC 3010, Australia}
\author{A.~K.~Romer}
\affiliation{Department of Physics and Astronomy, Pevensey Building, University of Sussex, Brighton, BN1 9QH, UK}
\author{S.~Samuroff}
\affiliation{Department of Physics, Northeastern University, Boston, MA 02115, USA}
\author{A.~Sarkar}
\affiliation{Kavli Institute for Astrophysics and Space Research, Massachusetts Institute of Technology, 70 Vassar St, Cambridge, MA 02139}
\author{E.~Sanchez}
\affiliation{Centro de Investigaciones Energ\'eticas, Medioambientales y Tecnol\'ogicas (CIEMAT), Madrid, Spain}
\author{I.~Sevilla-Noarbe}
\affiliation{Centro de Investigaciones Energ\'eticas, Medioambientales y Tecnol\'ogicas (CIEMAT), Madrid, Spain}
\author{M.~Smith}
\affiliation{Physics Department, Lancaster University, Lancaster, LA1 4YB, UK}
\author{E.~Suchyta}
\affiliation{Computer Science and Mathematics Division, Oak Ridge National Laboratory, Oak Ridge, TN 37831}
\author{M.~E.~C.~Swanson}
\affiliation{Center for Astrophysical Surveys, National Center for Supercomputing Applications, 1205 West Clark St., Urbana, IL 61801, USA}
\author{G.~Tarle}
\affiliation{Department of Physics, University of Michigan, Ann Arbor, MI 48109, USA}
\author{V.~Vikram}
\affiliation{Argone National Laboratory, 9700 S. Cass Avenue, Lemont, IL 60439, USA}
\author{N.~Weaverdyck}
\affiliation{Department of Astronomy, University of California, Berkeley,  501 Campbell Hall, Berkeley, CA 94720, USA}
\affiliation{Lawrence Berkeley National Laboratory, 1 Cyclotron Road, Berkeley, CA 94720, USA}
\author{J.~Weller}
\affiliation{
University Observatory, Faculty of Physics, Ludwig-Maximilians-Universit\"at, Scheinerstr.\ 1, 81679 Munich, Germany
}
\affiliation{Max Planck Institute for Extraterrestrial Physics, Giessenbachstr. 2, 85748 Garching, Germany}

\noaffiliation
\collaboration{the SPT and DES Collaborations}
\noaffiliation


\begin{abstract}
We present constraints on the $f(R)$ gravity model using a sample of 1,005 galaxy clusters in the redshift range $0.25 - 1.78$ that have been selected through the thermal Sunyaev-Zel'dovich effect (tSZE) from South Pole Telescope (SPT) data and subjected to optical and near-infrared confirmation with the Multi-component Matched Filter (MCMF) algorithm. We employ weak gravitational lensing mass calibration from the Dark Energy Survey (DES) Year 3 data for 688 clusters at $z < 0.95$ and from the Hubble Space Telescope (HST) for 39 clusters with $0.6 < z < 1.7$. Our cluster sample is a powerful probe of $f(R)$ gravity, because this model predicts a scale-dependent enhancement in the growth of structure, which impacts the halo mass function (HMF) at cluster mass scales. To account for these modified gravity effects on the HMF, our analysis employs a semi-analytical approach calibrated with numerical simulations. Combining calibrated cluster counts with primary cosmic microwave background (CMB) temperature and polarization anisotropy measurements from the Planck\,2018 release, we derive robust constraints on the $f(R)$ parameter $f_{R0}$. Our results, $\log_{10} |f_{R0}| < -5.32$ at the 95\,\% credible level, are the tightest current constraints on $f(R)$ gravity from cosmological scales. This upper limit rules out $f(R)$-like deviations from general relativity that result in more than a $\sim$20\% enhancement of the cluster population on mass scales $M_\mathrm{200c}>3\times10^{14}M_\odot$.
\end{abstract}

\maketitle


\section{\label{sec:intro}Introduction}
One of the most challenging questions in modern cosmology is understanding the nature of the accelerating expansion of the Universe \cite{Perlmutter1999,Riess1998}. Various cosmological theories have been proposed to explain this phenomenon. Within the framework of general relativity (GR), this acceleration can be explained by introducing a cosmological constant $\Lambda$ to the Einstein-Hilbert action, leading to the well-known $\Lambda$ cold dark matter ($\Lambda$CDM) model. 
However, adding a cosmological constant to the Einstein-Hilbert action offers little physical insight into the nature of Dark Energy. Therefore, there is strong motivation to consider modifications to the Einstein-Hilbert action that give rise to modified gravity models (see e.g., reviews \cite{JoyceEtal16,Koyama18,Baker19}). 

These modifications impact the growth of cosmic structures. Consequently, the abundance of massive galaxy clusters, as the end products of the hierarchical growth of cosmic structures, is sensitive to the different matter clustering and therefore, serves as an excellent probe for constraining modified gravity models and offering an independent test of GR.

In this work, we focus on a specific modified gravity model that introduces a non-linear function $f(R)$ of the scalar curvature $R$ into the Einstein-Hilbert action \cite{Buchdahl1970}. We employ the widely studied Hu \& Sawicki model for the function $f(R)$ \cite{Hu07}. Physically, this model introduces an additional gravitational-strength fifth force, altering structure formation in a scale-dependent manner and enhancing structure formation on galaxy cluster scales. The extent to which the $f(R)$ model deviates from GR is encoded in the single parameter $f_{R0}$, which has been constrained using various observations on cosmological scales.

Because the effects of $f(R)$ gravity persist to very small scales, constraints on galactic and solar-system scales are very stringent for this model. Studies using galaxy rotation curves and morphology report $\logfR < -6.1$ and $\logfR < -7.55$ at 95\% credible level, respectively \cite{Naik19,Desmond20}. However, because these studies probe small scales, systematics such as uncertainties in galaxy formation play a key role. This makes constraints from larger scales that are less sensitive to galaxy formation a complementary test. 

The current tightest constraint from clusters comes from a combination of ROSAT clusters, primary cosmic microwave background (CMB) anisotropy data, Supernovae (SNe), and baryonic acoustic oscillations (BAO), with an upper bound of $\logfR < -4.79$ at the 95\% credible level \cite{Cataneo16}. The recent analysis of eROSITA clusters reports $\logfR < -4.12$ at the 95\,\% credible level using clusters alone and marginalizing over the neutrino mass \cite{Artis24}. 

Stronger constraints from large scales probes are obtained from a weak-lensing peak analysis, which used data from the Canada-France-Hawaii-Telescope Lensing Survey (CFHTLenS), and are given by $\logfR < -5.16$ \cite{Xiangkun16}. Similar constraints come from the cross-correlation of galaxies with CMB lensing and galaxy weak-lensing, CMB, SNe, and BAO, reporting $\logfR < -4.61$ and $\logfR < -4.5$ at the 95\,\% credible level respectively \cite{Kou23,Hu16}. 

In the analysis presented here, we focus on the weak lensing informed galaxy cluster abundance to constrain $f(R)$ gravity. We are motivated to pursue this study partly because it has long been recognized that galaxy cluster surveys would be powerful probes of cosmic growth and therefore the action of gravity \citep{Wang98,Haiman01} and also because in recent years cluster surveys have been successfully employed to study the standard $\Lambda$CDM model \cite{Vikhlinin09,Benson13,Bocquet15,Dehaan16,Bocquet19,Abbott20,Chiu23,Bocquet24Ia,Bocquet24II,Ghirardini24} as well as modified gravity models \cite{Schmidt09b,Lombriser10,Cataneo14,Peirone17,Hagstotz19,Artis24}.

One of the most promising cluster samples currently available has been constructed using South Pole Telescope (SPT) \cite{Carlstrom11} survey data and the thermal Sunyaev-Zel’dovich effect (tSZE)~\cite{Sunyaev1972}. The tSZE is caused by high-energy electrons in the intra-cluster medium (ICM) scattering off CMB photons, resulting in a spectral distortion of the CMB at the cluster position. Because the tSZE is a direct tracer of the hot ICM, it enables the detection of massive galaxy clusters. Moreover, the cluster tSZE signature is strongly mass dependent and approximately redshift-independent and thus can be employed to identify galaxy clusters up to the highest redshifts where clusters of sufficient mass exist. To constrain cosmological parameters with galaxy clusters, one has to relate observables such as the tSZE detection significance to the underlying halo mass at all relevant redshifts. These observable--mass relations can be empirically calibrated using weak gravitational lensing data and are typically modeled as power laws in cluster mass and redshift. 

In this study, we employ the sample of 1,005 galaxy clusters detected using SPT data and confirmed using the MCMF algorithm \citep{Klein18,Klein24,Bleem24} with optical and near-infrared data from the Dark Energy Survey (DES) \citep{flaugher15,DES16,DES18DR1} and the Wide-field Infrared Survey Explorer (WISE) \cite{WISEobservatory}. To obtain mass estimates for the cluster sample, we use weak-lensing measurements from DES and targeted observations from the Hubble Space Telescope (HST). The $f(R)$ analysis framework employed for the SPT clusters with mass calibration from DES and HST is based on the state-of-the-art method developed for the recent $\Lambda$CDM analyses of this same sample \cite{Bocquet24Ia,Bocquet24II} (hereafter \citetalias{Bocquet24Ia} and \citetalias{Bocquet24II}). In a recent paper, this same framework was modified and employed to carry out validation tests and forecasts for $f(R)$ gravity constraints from upcoming Stage-III and Stage-IV surveys \cite{Vogt24} (hereafter \citetalias{Vogt24}).

Following \citetalias{Vogt24}, we incorporate $f(R)$ gravity into our analysis by modifying the halo mass function (HMF), which is enhanced relative to GR, in a mass- and redshift-dependent way. 
We implement this modification by introducing a multiplicative factor
to the GR HMF \cite{Shandera13,Cataneo14}, dependent on the spherical collapse threshold for halo collapse in $f(R)$ gravity for which we use a semi-analytical model \cite{Li12a,Lombriser13}. In the present analysis we calibrate this semi-analytical HMF model against the $f(R)$ FORGE numerical simulations to obtain a more accurate halo mass function \cite{Arnold21}. The simulations are not used directly to predict the HMF, because the available mass range from the simulations is limited, whereas the HMF from the semi-analytical model can be calculated for the wide mass range needed in this analysis.

This paper is organized as follows. Section~\ref{sec:data} presents a summary of the SPT cluster dataset and the DES and HST weak-lensing data. We review in Sec.~\ref{sec:MG} $f(R)$ gravity and the $f(R)$ HMF model used in this work as well as the calibration to the FORGE simulations. In Sec.~\ref{sec:analysis}, we discuss our analysis method, including the observable--mass relations, weak-lensing model, likelihood approach, and priors. The results are presented in Sec.~\ref{sec:results}, and we conclude with a summary in Sec.~\ref{sec:summary}.

Throughout this paper, $\mathcal{U}(a, b)$ denotes a uniform distribution between limits $a$ and $b$, and $\mathcal{N}(\mu, \sigma^2)$ is a Gaussian distribution with mean $\mu$ and variance $\sigma^2$. We adopt the halo mass definition $M_{200\mathrm{c}}$, which is the mass within a radius where the mean density is 200 times the critical density.

\section{\label{sec:data}Data}
This section gives a brief summary of the cluster and weak-lensing data we use in this work. A detailed description of the data products is presented in \citetalias{Bocquet24Ia}.

\subsection{\label{subsec:SPT} SPT cluster catalog}
The tSZE selected cluster catalogs from the SPT-SZ, SPTpol~ECS and SPTpol~500d surveys employed here cover a total solid angle of $5,270\,\rm deg^2$ of the southern sky \citep{Bleem15,Bleem20,Klein24,Bleem24}. Note that the whole SPTpol~500d survey lies within the SPT-SZ footprint, and we use only the data from the deeper SPTpol~500d survey in the overlapping region. Cluster candidates of these surveys are selected in tSZE detection significance \zetahat \ and confirmed using optical and infrared data, which also add redshift information.

Over the whole SPT survey region, only clusters with $z > 0.25$ are included in the sample, because 
at low redshift the selection function becomes harder to model due to the impact of the filtering of atmospheric and primary CMB signals. The angular size of clusters becomes larger at low redshift, and therefore a larger fraction of the cluster signal to lost due to this filtering.

In the SPT survey region that is not covered by DES ($1,327\,\rm deg^2$, ca.\ $27\,\%$ of the total solid angle), the cluster candidates are confirmed by targeted observations that also provide redshift measurements. Candidates in this region are selected by 
    \begin{equation}
      \label{eq:select_noDES}
      \begin{split}
        \zetahat&>5 \\
        z&>0.25 \,,
      \end{split}
    \end{equation}
which results in a sample with 110 clusters and a purity $\gtrsim 95\,\%$ \citep{Bleem15,Bleem20}.

For the region covered by DES ($3,567\,\rm deg^2$, $75\,\%$ of the SPT area) cluster candidates are optically confirmed using the Multi-Component Matched Filter cluster confirmation tool \citep[MCMF;][]{Klein18,Klein24}, and we follow the work of Refs.~\cite{Bleem24,Klein24}. Moreover, measurements for cluster redshift $z$, optical richness \lambdahat\ and optical center position are obtained using MCMF. Because DES data are only reliable for redshifts $z \leq 1.1$, WISE data are used to compute richnesses and redshifts for clusters with $z > 1.1$. There is the chance that an overdensity of galaxies is a random superposition along the line of sight of a tSZE noise fluctuation. 
To exclude chance associations, we use a redshift dependent richness cut, $\hat \lambda_{\rm min}(z)$, provided by the MCMF tool, to ensure a constant purity $>98\,\%$ over the entire redshift range.
A cluster candidate is then confirmed if the measured richness is larger than this threshold. 

Using MCMF for cluster confirmation allows us to validate clusters with lower tSZE detection significance while achieving a high sample purity, and results in a cluster sample, which is
at least
$30\,\%$ larger than a solely tSZE selected sample with the same purity.

Due to the different depths of the individual tSZE surveys, different selection thresholds in \zetahat\ are applied to obtain an approximately constant purity for the combined SPT sample. The selection criteria are
    \begin{equation}
      \label{eq:select_DES}
      \begin{split}
        \zetahat&>4.25 \,/\, 4.5 \,/\, 5 \,\,(\text{500d / SZ / ECS})\,, \\
        \lambdahat&>\hat \lambda_\mathrm{min}(z) \,, \\
        z&>0.25\,.
      \end{split}
    \end{equation}
These selections result in a sample of 895 confirmed clusters over this region.

To summarize, the total cluster sample consists of 1,005 clusters, each characterized by the observables tSZE detection significance $\zetahat$ and redshift $z$. In the SPT area covered by DES, these clusters also have additional measurements of richness $\lambdahat$ and cluster center position from either DES ($z \leq 1.1$) or WISE ($z > 1.1$). 

\begin{figure*}
        \centering
        \includegraphics[width=\textwidth]{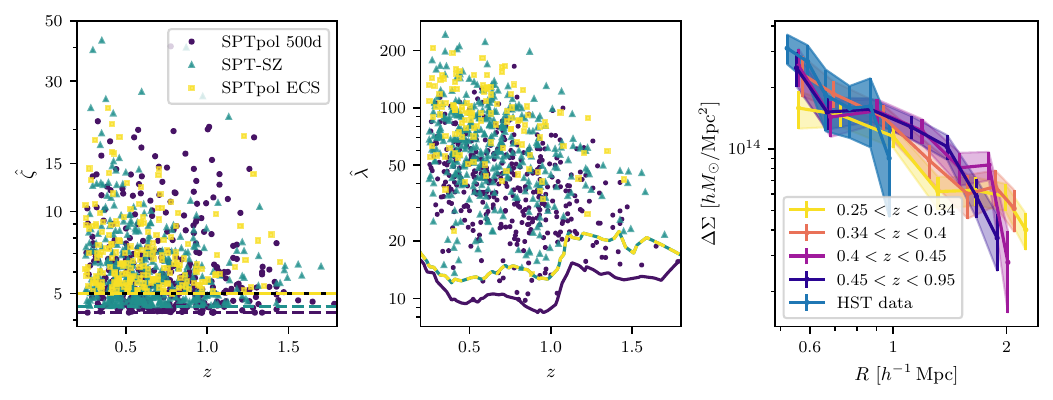}
        \vskip-0.25cm
        \caption{
        \textit{Left}: tSZE detection significance $\zetahat$ and redshift distribution for the three SPT surveys. Dashed-colored lines show the detection threshold for the corresponding survey in the region that overlaps with DES. The black dashed line shows the $\zetahat$ threshold for the clusters outside of the DES region (same as the cut for the SPTpol~ECS survey). 
        \textit{Middle}: optical richness $\lambdahat$ and redshift distribution of the cluster sample, color coded by survey. Colored lines correspond to the $\lambdahat_\mathrm{min}(z)$ detection threshold of the given survey. The SPTpol~500d is significantly deeper than the other two surveys (yellow-green dashed line) and thus a lower $\lambdahat(z)$ threshold is applied (solid purple line). 
        \textit{Right}: averaged weak-lensing inferred projected matter profiles from DES~Y3 data shown for four redshift bins in purple to yellow. The redshift bins are chosen such that the signal-to-noise is approximately equal in each bin. A similar profile derived from HST data are shown in blue. Over the radial range used for weak-lensing mass calibration, the DES~Y3 data have a signal-to-noise of $31.2$ compared to the HST data with $9.7$.}
        \label{fig:data}
    \end{figure*}

Figure~\ref{fig:data} shows the 1,005 confirmed cluster sample in the space of tSZE detection significance $\zetahat$ and redshift (left) as well as in optical richness $\lambdahat$ and redshift (middle). The clusters are color coded according to which of the three SPT surveys they originate from. In addition, the corresponding selection thresholds are shown as lines in each observable. One can see in the $\zetahat$--$z$ distribution that while the sample extends to $z\sim1.8$ the bulk of the clusters lie at $z<1$. In addition, this distribution makes clear that the majority of the cluster lies close to the tSZE detection thresholds of the surveys. The $\lambdahat$--$z$ distribution displayed in the middle panel shows that the bulk of the clusters have relatively high richnesses $\lambdahat>30$, and that the optical selection thresholds are not significantly impacting the completeness of the confirmed cluster sample. Indeed, the MCMF selection thresholds largely impact the high purity of the sample, because the noise fluctuations present in the tSZE selected candidate clusters are efficiently removed due to their low optical richnesses ($\lambdahat<10$).
%

\subsection{\label{subsec:DES} DES Y3 weak-lensing data}
The DES was conducted in the $griz$Y bands and covers a sky area of $5,000\,\rm deg^2$.
The DES~Y3 weak lensing shape catalog \cite{Gatti22} utilized data from the first three years of observations and covers approximately $4,143\,\rm deg^2$ of the sky after masking. $3,567\,\rm deg^2$ of the DES region overlaps with the SPT surveys, corresponding to $75\,\%$ of the whole SPT survey area. The weak-lensing shape catalog of DES~Y3 is built with the \textsc{Metacalibration} pipeline from the $r$, $i$ and $z$ bands \cite{Huff17,Sheldon17}. Lensing source galaxies are selected in four tomographic redshift bins as employed in the $3\times2$pt analysis of DES \cite{DES_Y3_3x2pt}. 

For each cluster in the overlapping region of SPT and DES, we use the weak-lensing shear profiles within the radial range $0.5<r/(h^{-1}\mathrm{Mpc})<3.2\, (1+z_\mathrm{cluster})^{-1}$\footnote{These regions are calculated in a fiducial cosmology with $\Omega_{\rm m} = 0.3$ and $h = 0.7$} around the optical cluster center. The lower limit on the radial range excludes the inner region of the cluster, which is largely affected by feedback from active galactic nuclei, miscentering, blending, cluster member contamination and non-linear shear. The upper limit on the radial range guarantees that only the one-halo term region is used for mass calibration \cite{Grandis21}. We only use weak-lensing data from DES for clusters with $z < 0.95$, corresponding to the median redshift of the highest redshift tomographic bin \citepalias{Bocquet24Ia,Bocquet24II}. 

In our analysis, we account for systematic and statistical uncertainties such as cluster member contamination, miscentering of the shear profile, shear and photo-z calibration, halo mass modeling and the impact of large-scale structure. A detailed description of the modeling of these uncertainties can be found in detail in \citetalias{Bocquet24Ia}, Section~V. 
Note that the calibration of these uncertainties was performed within the $\Lambda$CDM paradigm, but we expect these to not change significantly in $f(R)$ gravity \cite{Ruan24}.
In total, our analysis includes 688 cluster shear profiles from $555,912$ source galaxies with an average of 808 shear source galaxies per cluster \citepalias{Bocquet24Ia}. We show for illustration the averaged matter density profile of the DES~Y3 data in the right panel of Fig.~\ref{fig:data}. The profiles are broken into four redshift ranges of comparable signal-to-noise. The combined dataset corresponds to a $31.2\,\sigma$ detection of the matter profiles of these clusters.

\subsection{\label{subsec:HST} HST weak-lensing data}
DES lensing data are only reliable for $z \lesssim 0.95$ and therefore, we complement the weak-lensing dataset with HST data to obtain weak-lensing information for the mass calibration at high redshift. We use the HST-39 dataset \cite{Schrabback18,Schrabback21,Zohren22} to obtain weak-lensing shear profiles. This dataset contains 39 clusters of our tSZE selected sample in the redshift range $0.6 - 1.7$. More details about the dataset and the analysis can be found in Refs.~\cite{Schrabback18,Schrabback21,Raihan20,Hernandes20,Zohren22,Sommer22}. The averaged matter density profile from the 39 clusters from HST is shown in Fig.~\ref{fig:data} in the right panel in blue. These data correspond to a $9.7\,\sigma$ detection of the matter profiles of these halos.

\section{\label{sec:MG}\texorpdfstring{$f(R)$}{f(R)} Modified gravity}
$f(R)$ gravity modifies GR by introducing an arbitrary function $f(R)$ of the Ricci scalar $R$ into the Einstein-Hilbert action \cite{Buchdahl1970}
    \begin{equation}
        \label{eq:fR_EH_action}
        S = \int \dd^4 x \sqrt{-g} \left [ \frac{R + f(R)}{16\pi G} + \mathcal{L}_m \right ] \, ,
    \end{equation}
where $g$ denotes the determinant of the metric tensor, $G$ is the gravitational constant and $\mathcal{L}_m$ is the matter Lagrangian density. Note that we use natural units where $c = \hbar = 1$ and if $f(R)= -2\Lambda$ we recover GR plus a cosmological constant, hence a $\Lambda$CDM cosmology. The field equation obtained from this modified action takes the form
    \begin{equation}
        \label{eq:fR_field_eq}
        \begin{split}
            G_{\mu \nu} + f_R R_{\mu \nu} -& \left (\frac{f}{2} - \square f_R \right ) g_{\mu \nu} \\
            & \hspace{1.5cm} - \nabla_\mu \nabla_\nu f_R  = 8 \pi G T_{\mu \nu} \, .
        \end{split}
    \end{equation}
Here $G_{\mu\nu}$ is the Einstein tensor, $R_{\mu\nu}$ is the Ricci tensor, $T_{\mu\nu}$ represents the energy-momentum tensor and $f_R = \dd f(R) / \dd R$ is an additional scalar degree of freedom, which indicates the strength of the modifications to GR. 

The equation of motion for $f_R$ is derived from the trace of the field equation in the quasistatic and weak-field limit
    \begin{equation}
        \label{eq:EoM_fR}
        \nabla^2 \delta f_R = \frac{1}{3} (\delta R - 8 \pi G \delta \rho) \, ,
    \end{equation}
where $\delta x = x - \bar x $ is the perturbation of the quantity $x$ with respect to the cosmic mean. Additionally, from the time-time component of the field equation, Eq~\eqref{eq:fR_field_eq}, we derive the modified Poisson equation in $f(R)$ gravity:
    \begin{equation}
        \label{eq:poisson_eq}
        \nabla^2 \Phi = \frac{16 \pi G}{3} \delta \rho - \frac{1}{6} \delta R \,,
    \end{equation}
with $\Phi$ the Newtonian potential which is defined via $2\Phi = \delta g_{00}/g_{00}$. Combining Eqs.~\eqref{eq:EoM_fR} and \eqref{eq:poisson_eq} shows how the Poisson equation and thus the structure growth depends on the strength of the $f(R)$ gravity model
    \begin{equation}
        \label{eq:poisson_eq_2}
        \nabla^2 \Phi = 4 \pi G \delta \rho - \frac{1}{2} \nabla^2 \delta f_R \,.
    \end{equation}
Compared to GR the Poisson equation includes a term proportional to the Laplacian of $\delta f_R$ and thus depends directly on the strength of the model.

The modified Poisson equation, Eq.~\eqref{eq:poisson_eq}, shows that the strength of the gravitational force depends on the environment. In unscreened (``low curvature'') regions when $|f_R| \gg |\Phi|$ it follows that $\delta R \ll 8\pi G \delta \rho$ and consequently the $f(R)$ Poisson equation corresponds to a modified Poisson equation with gravitational forces enhanced by a factor of $4/3$ compared to GR. In screened (``high curvature'') environments, \ie $|f_R| \ll |\Phi|$, the Ricci scalar $R$ is approximately $ 8\pi G \delta \rho$ and thus Eq.~\eqref{eq:poisson_eq} reduces to the unmodified GR Poisson equation: in other words, in high-density regions $f(R)$ gravity falls back to GR \cite[see][for a detailed discussion]{Cataneo16,Hu07}. 
This property of driving $f_R \to 0$ in high-density regions is the so-called chameleon screening mechanism \cite{Khoury04}, which makes $f(R)$ gravity consistent with Solar systems tests \cite{Hu07,Burrage18,Fischer24}. The two limits show that structure growth in $f(R)$ gravity is environment-dependent in contrast to GR.

In this paper, we adopt the widely used and studied Hu \& Sawicki model \cite{Hu07}
    \begin{equation}
        \label{eq:fR_Hu07}
        f(R) = - m^2 \frac{c_1\left( \frac{R}{m^2} \right)^n}{c_2\left( \frac{R}{m^2} \right)^n + 1} \, ,
    \end{equation}
with $m^2 = \Omega_\mathrm{m} H^2_0$, $H_0$ the Hubble constant and the free parameters $n$, $c_1$, $c_2$. In the high curvature regime we have $c_2^{1/n}R/m^2 \gg 1$ and Eq.~\eqref{eq:fR_Hu07} is approximately given by
    \begin{equation}
        \label{eq:fR_approx_high_curv}
        f(R) \approx - m^2 \frac{c_1}{c_2} - \frac{f_{R0} R_0^{n+1}}{nR^n} \, .
    \end{equation}
Here $R_0$ is the present background Ricci scalar and $f_{R0} := f_R(R_0)$ is the parameter that quantifies the strength of this $f(R)$ gravity model.

To have a modified gravity scenario that is active at late times and large scales, when the acceleration of the Universe happens, and that does not spoil the successful description of early Universe observables such as BBN and CMB the function $f(R)$ has to satisfy $f_R < 0$, \ie it is a decreasing function of $R$. Therefore, the approximation above is correct up to order $\sim (f_{R0})^2$, and since the constraints from our cluster sample are better than $|f_{R0}| \sim 10^{-4}$, the approximation in Eq.~\eqref{eq:fR_approx_high_curv} is entirely sufficient. We work with \logfR\ for numerical convenience, and, given that the theory gives no strong prior on the scale of $f_{R0}$, we impose a uniform prior on \logfR. 

To obtain an expansion history consistent with $\Lambda$CDM in the limit $|f_{R0}| \to 0$, the parameters $c_1$ and $c_2$ are given by
    \begin{equation}
        \label{eq:condi_params_fR}
        \frac{c_1}{c_2} = 6 \frac{\Omega_\Lambda }{\Omega_{\mathrm{m}} } \, .
    \end{equation}

We further adopt $n = 1$ for the Hu \& Sawicki model; see Ref.~\cite{Ferraro2011} for an approach to approximately rescale constraints from $n=1$ to other values of $n$.

\subsection{\label{subsec:HMF} The halo mass function in \texorpdfstring{$f(R)$}{f(R)} gravity}
As we constrain $f(R)$ gravity with the help of cluster abundance datasets we need a model for the distribution of halo mass and redshift within this theory, in other words the (differential) halo mass function. In this work we adopt a two-component model \cite{Shandera13,Cataneo14} for which the first component is the HMF in GR, and the second factor models the enhancement of the $f(R)$ gravity, which accounts for the scale-dependent clustering due to the scale-dependent structure growth described in Sec.~\ref{sec:MG}:
    \begin{equation}
        \label{eq:fR_HMF}
        \frac{\dd n}{\dd \mathrm{ln} M} = \left. \frac{\dd n}{\dd \mathrm{ln} M} \right\vert_{\mathrm{GR}} \times \mathcal{R}\, .
    \end{equation}
For the GR HMF, we use the halo mass function from Ref.~\cite{Tinker08}
    \begin{equation}
        \label{eq:Tinker_HMF}
        \left. \frac{\dd n}{\dd \mathrm{ln} M} \right\vert_{\mathrm{GR}} = - \frac{\bar \rho_{\mathrm{m}}}{2M} f(\sigma)_{\mathrm{T}} \frac{\dd \mathrm{ln} \sigma^2}{\dd \mathrm{ln} M}\, ,
    \end{equation}
with the multiplicity function of the form
    \begin{equation}
        \label{eq:Tinker_multi_fct}
        f(\sigma)_{\mathrm{T}} = \tilde A \left [ \left(\frac{\sigma}{\tilde b}\right)^{-\tilde a} +1 \right]e^{-\frac{\tilde c}{\sigma^2}}\, ,
    \end{equation}
where $\tilde A,\ \tilde a,\ \tilde b$ and $\tilde c$ are parameters calibrated using N-body simulations~\cite[see][table 2]{Tinker08} and $\sigma=\sigma(M)$ is the variance of the overdensity field on a mass scale $M$ in the corresponding GR cosmology.

The enhancement factor $\mathcal{R}$ is calculated from the ratio of the Sheth-Tormen HMF \cite{Sheth1999} in $f(R)$ gravity to GR, \ie 
    \begin{equation}
        \label{eq:ratio_ST_HMF}
        \mathcal{R} = \frac{\left. \frac{\dd n}{\dd \mathrm{ln} M} \right\vert_{\mathrm{ST},\, f(R)}}{\left. \frac{\dd n}{\dd \mathrm{ln} M} \right\vert_{\mathrm{ST},\, \mathrm{GR}}}  \, .
    \end{equation}
This HMF can account for the scale-dependent clustering through the spherical collapse threshold $\delta_{\rm crit}$. In this work, we calculate the spherical collapse threshold $\delta_{\rm crit}$ from the spherical collapse model of Refs.~\cite{Li12a,Lombriser13}. With this, the Sheth-Tormen HMF is given by 
    \begin{equation}
        \label{eq:ST_HMF}
        \left. \frac{\dd n}{\dd \mathrm{ln} M} \right\vert_{\mathrm{ST}} =  \frac{\bar \rho_{\mathrm{m}}}{M} f(\nu)_{\mathrm{ST}} \left[ \frac{\dd\mathrm{ln} \delta_{\rm crit}}{\dd\mathrm{ln} M} - \frac{1}{2} \frac{\dd\mathrm{ln} \sigma^2}{\dd\mathrm{ln} M} \right ]  \, .
    \end{equation}
Here $\nu = \delta_{\rm crit}/ \sigma$ is the peak height and $f(\nu)_{\mathrm{ST}}$ refers to the Sheth-Tormen multiplicity function \cite{Sheth1999},
    \begin{equation}
        \label{eq:ST_multi_fct}
        f(\nu)_{\mathrm{ST}} = A \sqrt{\frac{a\nu^2}{2\pi}} \left[1+(a\nu^2)^{-p}\right]e^{-\frac{a\nu^2}{2}}\, ,
    \end{equation}
with $A,\ a$, $p$ are free parameters for which we adopt the parameters of Ref.~\cite{Despali16}.

The change in the clustering and thus in the HMF is completely captured by the scale-dependent spherical collapse threshold $\delta_{\rm crit}$ and therefore, we use an effective variance $\sigma$ derived from the corresponding GR cosmology \cite{Lombriser13,Lombriser14}. 
Moreover, when including massive neutrinos in a GR cosmology, it has been shown that the semi-analytical HMF shape is closer to universal if the halo mass and variance are calculated neglecting the neutrino component \cite{Ichiki12,Costanzi13}, \ie, using the baryon and cold dark matter power spectrum instead of the total matter power spectrum for the variance. Because there is a degeneracy between massive neutrinos and \logfR\ (at least for high values of $\Sigma m_\nu$ and \logfR) \citep[see \eg][]{Motohashi13,Baldi14,Wright19} we have to account for massive neutrinos in our analysis and as we expect the effects of massive neutrinos to be approximately the same in GR and $f(R)$ gravity we use the approach of Refs.~\cite{Ichiki12,Costanzi13} to account for the nonzero total neutrino mass. 

The derivation of $\delta_{\rm crit}$ with the spherical collapse model requires solving a computationally expensive system of coupled differential equations for each value of mass, redshift and cosmology ~\cite{Li12a,Lombriser13}.
To speed up the calculations we use emulators to predict the values of $\delta_{\rm crit}$ and $\dd \delta_{\rm crit} / \dd \mathrm{ln} M$. The emulators adopted in this analysis are presented in \citetalias{Vogt24}.

\begin{figure}
        \centering
        \includegraphics[width=\linewidth]{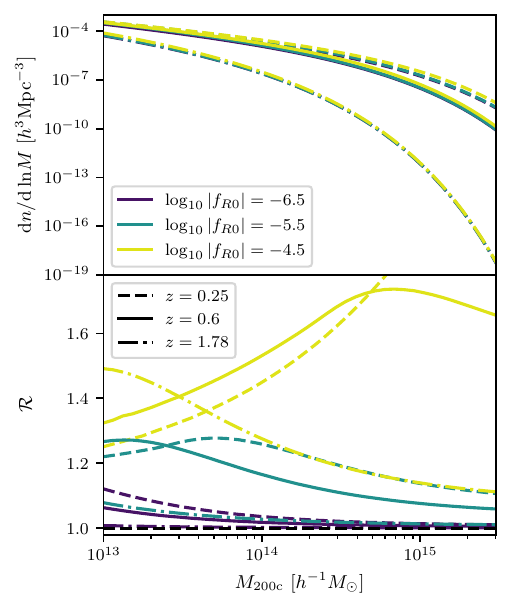}
        \vskip-0.25cm
        \caption{
        The $f(R)$ HMF, Eq.~\eqref{eq:fR_HMF}, (top) and the ratio $\mathcal{R}$ of the $f(R)$ and the GR Sheth-Tormen HMF, Eq.~\eqref{eq:ST_HMF}, (bottom) for different values of $\log_{10}|f_{R0}|$ (different colors) and three redshifts (different line styles) corresponding to the minimum, mean and maximum redshift of the SPT cluster sample. 
        The deviation from GR depends on the strength of the $f(R)$ gravity and redshift, whereby weaker $f(R)$ models and higher redshifts show less enhancement in the HMF relative to GR.
        }
        \label{fig:HMF_fR}
    \end{figure}

Figure~\ref{fig:HMF_fR} shows the $f(R)$ HMF on top and the ratio $\mathcal{R}$ on the bottom for different values of \logfR\ and redshift. In $f(R)$ gravity, the HMF is larger compared to GR as expected from the enhanced structure growth. The shape and strength of the enhancement depend on the value of \logfR\ with a larger enhancement for stronger $f(R)$ models. Furthermore, the enhancement becomes smaller at higher redshifts, because modified gravity falls back to GR in the early Universe. 
A large enhancement of the most massive halos is only seen for the strongest model with $\logfR = -4.5$ and at lower redshifts, because even the most massive halos are only partially screened in this scenario. In the other cases, the potential $\Phi$ of the high mass halos becomes compatible with $f_R$ and the fifth force is screened in these massive halos, reducing the modified gravity effect on the halo abundance.
%

\subsection{\label{subsec:HMF_cali} HMF calibration using FORGE simulations}
The model for the HMF presented above follows a semi-analytical approach where the $f(R)$ gravity is incorporated via the spherical collapse threshold $\delta_{\rm crit}$ calculated from spherical collapse theory in $f(R)$ gravity. Besides semi-analytical models, the HMF can also be derived from simulations. Therefore, we make a comparison with $f(R)$ simulations to validate the semi-analytical model and to check for any kind of discrepancy.

In this work, we use the state-of-the-art FORGE $N$-body simulations which encompass 49 $f(R)$ gravity cosmologies (nodes) \cite{Arnold21}. The simulations sample the cosmological parameters $\Omega_m$, $h$, $S_8^{\rm GR} = \sigma_8^{\rm GR} \sqrt{\Omega_{\rm m}/0.3}$, and $\logfR$ (hereafter FORGE parameters) with a latin hypercube, while the other cosmological parameters are fixed to $n_s = 0.9652$ $\Omega_{\rm b} = 0.049199$, $\Omega_\nu = 0$ and $\Omega_{\Lambda} = 1 - \Omega_{\rm m}$ \cite{Arnold21}. Each FORGE $f(R)$ gravity node has a $\Lambda$CDM counterpart. The parameter ranges explored in the FORGE simulations are
    \begin{alignat}{4}
      \label{eq:FORGE_params}
        0.11 &< \qquad \Omega_{\rm m} &&< 0.54 \, , \nonumber \\ 
        0.61 &< \qquad h &&< 0.81 \, ,  \nonumber \\
        0.6 &< \qquad S_8^{\rm GR} &&< 0.9 \, , \\
        -6.17 &< \logfR &&< -4.51 \,. \nonumber
    \end{alignat}
The ranges in $\Omega_{\rm m}$ and $S_8^{\rm GR}$ translate to a range in $\sigma_8^{\rm GR}$ of $0.49 < \sigma_8^{\rm GR} < 1.31$. 

Reference~\cite{Ruan24} presents an emulator for the $f(R)$ HMF that is based upon the FORGE simulations. 
In this work, the authors trained a neural network to directly predict the enhancement factor of the HMF due to $f(R)$ gravity for each mass, redshift and set of FORGE parameters. 
However, by design, the resulting halo mass function is only valid within the mass, redshift and cosmological parameter range used in the neural network training set. 
To make predictions outside the mass range available within the FORGE simulations, we calibrate the semi-analytical HMF model, which is valid up to halo masses of $10^{16}\,h^{-1}\,M_\odot$, with the HMF retrieved from the FORGE simulations.

To calibrate the semi-analytical HMF model with the FORGE simulations we use the 
high-resolution simulations with $1024^3$ dark matter particles in a box with length $L = 500\,h^{-1}\,\rm Mpc$ and a mass resolution of $9.5\times 10^9 (\Omega_{\rm m} / 0.3)\, h^{-1}M_{\odot}$. 
We extract halo catalogs from all $f(R)$ simulations as well as from their corresponding $\Lambda$CDM nodes at redshifts $z = 0.00, 0.25, 0.5, 0.75, 1.00, 1.25, 1.50, 1.75$, and $2.00$, using a bin width of 0.1 in $\log_{10}M$ within the mass range $10^{13}\,h^{-1}M_\odot \leq M_{200\rm c} \leq 5\times10^{15}\,h^{-1}M_\odot$. To ensure no empty bins at the high-mass end, we combine the last high-mass bins into one (large) bin such that it contains at least 20 halos \footnote{with this approach the size of the last bin can vary because the only requirement is that the last bin contains at least 20 halos.}. We compute covariance matrices for each halo catalog using $5^3$ jackknife samples to account for noise due to sample variance and shot noise.

We characterize the difference between the semi-analytical model and the simulation by comparing the enhancement in the HMF, $\mathcal{R}$, from the simulations and the semi-analytical HMF model. This has the advantage that the cosmic variance in the simulations partially cancels in this ratio. 
With the halo catalogs from $f(R)$ gravity and $\Lambda$CDM FORGE simulations, we can calculate the enhancement in the cluster counts, $\mathcal{R}_{\rm FORGE}$, for each node and redshift. The enhancement from the semi-analytical model, $\mathcal{R}_{\rm SAM}$, is also calculated for each FORGE cosmology and redshift. 

\begin{figure*}
        \centering
        \includegraphics[width=\textwidth]{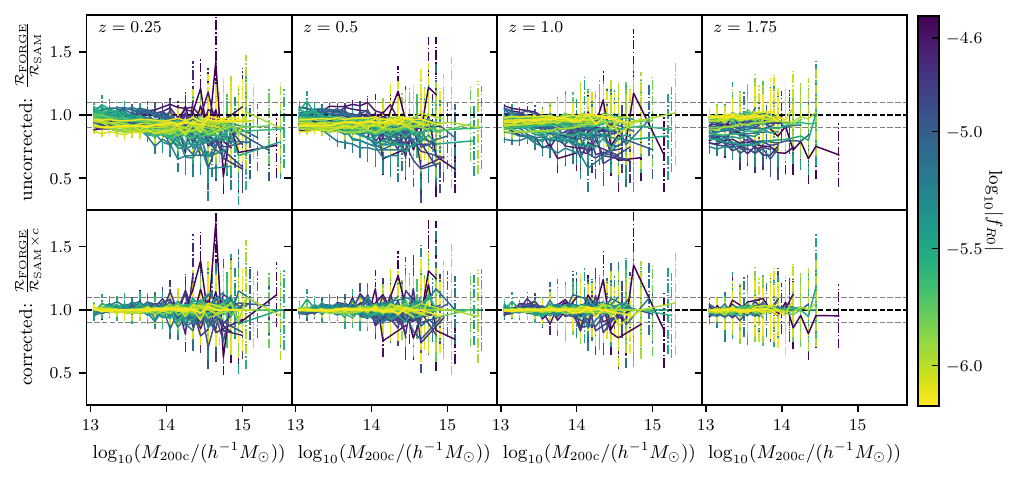}
        \vskip-0.25cm
        \caption{Comparison of the enhancement in the HMF from the FORGE simulations, $\mathcal{R}_{\rm FORGE}$, and the semi-analytical HMF model, $\mathcal{R}_{\rm SAM}$ for different redshifts, color-coded by the \logfR\ values. Gray dashed lines are plotted at $10\,\%$ deviation to guide the eye. The upper four plots show the comparison of the FORGE simulations to the semi-analytical model, Eq.~\eqref{eq:fR_HMF}, without a correction.
        The four plots at the bottom show the comparison to the corrected semi-analytical HMF, Eq.~\eqref{eq:fR_HMF_FORGE}, leading to a better agreement between the two HMF enhancements. Error bars are derived from the jackknife covariance of the FORGE simulations.
        }
        \label{fig:comp_FORGE}
    \end{figure*}

The top panels of Fig.~\ref{fig:comp_FORGE} show the ratios between the enhancements in each mass bin, \ie $\mathcal{R}_{\rm FORGE} / \mathcal{R}_{\rm SAM}$, for four of the nine different redshifts with error bars derived from the jackknife covariance of the simulations. There is a bias between the semi-analytical HMF and the FORGE simulations, which varies with mass, redshift and the FORGE parameters. In Fig.~\ref{fig:comp_FORGE} the ratio is color-coded based on the value of \logfR\ and the discrepancy is larger with higher values of \logfR. Note that also the other cosmological parameters can drive the difference between the semi-analytical HMF and the simulations. Therefore, we assume that the bias depends on all FORGE parameters. Overall the semi-analytical HMF model predicts more clusters than the FORGE simulations, and the agreement is better for weaker $f(R)$ models, \ie smaller values of \logfR.

Based on the top panels of Fig.~\ref{fig:comp_FORGE} we model the ratio and thus the correction to the semi-analytical HMF with a broken linear function in $\log_{10} M$ with a pivot scale $\log_{10}M_{\rm piv}$ and smooth transition with strength $k$ between the two linear functions at given logarithmic mass $\log_{10} M_1$. To be precise, for $\log_{10} M$ smaller than $\log_{10} M_1$, the ratio is modeled with a linear function in $\log_{10} M$ with slope $a$ and intercept $b$. For $\log_{10} M$ larger than $\log_{10} M_1$, the same ratio is modeled by a different linear function with slope $a_2 = a + \Delta a$ and intercept fixed 
by continuity at the transition mass $M_1$. We then smooth the transition between the two linear relations by a power-law interpolation controlled by a parameter $k$. For fixed pivot mass, transition mass and smoothing strength, the fitting function for the correction has three free parameters and is given by
    \begin{small}
    \begin{equation}
      \label{eq:double_lin_corr}
      \begin{split}
        c(M, \boldsymbol{p}, z) = &a(\boldsymbol{p}, z) (\log_{10}M - \log_{10}M_{\rm piv}) + b(\boldsymbol{p}, z) \\
        &+ \Delta a (\boldsymbol{p}, z) \frac{  \ln(1+e^{-k(\log_{10}M - \log_{10}M_1)})}{k}  \\
        &  + \Delta a (\boldsymbol{p}, z) (\log_{10}M - \log_{10}M_1)\, .
      \end{split}
    \end{equation}
    \end{small}
Here $\boldsymbol{p}$ represents the vector containing the FORGE parameters, $a(\boldsymbol{p}, z)$, $\Delta a(\boldsymbol{p}, z),\ b(\boldsymbol{p}, z)$ are the fitting parameters of the correction for which we assume a dependence on the FORGE parameters and redshift. We decided to fit for the parameter $\Delta a(\boldsymbol{p}, z)$ instead of $a_2(\boldsymbol{p}, z) = a(\boldsymbol{p}, z) + \Delta a(\boldsymbol{p}, z)$ because it showed a better behavior during fitting. 

The parameters of the broken linear function are obtained by fitting the correction function, Eq.~\eqref{eq:double_lin_corr}, to the ratio between the enhancements, $\mathcal{R}_{\rm FORGE} / {R}_{\rm SAM}$, where we fix the parameters $k = 2$, $\log_{10} M_1 = 14.5$ and $\log_{10} M_{\rm piv} = 13$. With this correction, we achieve good agreement with the simulations within the error bars as shown in the bottom panels of Figure~\ref{fig:comp_FORGE}.

Because we assume that the fitting parameters depend on cosmology and redshift we have to predict these parameters for an arbitrary cosmology and redshift to use the calibrated HMF in our analysis. To do so we build emulators based on Gaussian process regression to predict the three fitting parameters as a function of the FORGE parameters and redshift.

With the emulators, we can calculate the correction to the HMF for an arbitrary cosmology (within the parameter ranges Eq.~\eqref{eq:FORGE_params}), and our calibrated semi-analytical HMF is given by
    \begin{equation}
        \label{eq:fR_HMF_FORGE}
        \frac{\dd n}{\dd \mathrm{ln} M}  = \left. \frac{\dd n}{\dd \mathrm{ln} M} \right\vert_{\mathrm{GR}} \times  c \times\mathcal{R} \, ,
    \end{equation}
where $\mathcal{R}$ is the enhancement factor in the semi-analytical halo mass function from Eq.~\eqref{eq:ratio_ST_HMF} and $c$ is the correction function, Eq.~\eqref{eq:double_lin_corr}, calibrated with the FORGE simulations. 

The semi-analytical HMF model always predicts an enhancement and also the FORGE simulations show no sign of suppression apart from noise at the high mass end. Due to the presence of this noise, and because we model the correction function $c(M, \boldsymbol{p}, z)$ with a broken linear function, the calibrated enhancement of the HMF, $c\times\mathcal{R}$, may be reduced to values below unity at high halo masses. In this case, we set $c\times\mathcal{R}=1$ for all masses above the mass where the corrected HMF factor drops below unity to be consistent with the theory and the simulations.

Using the 49 FORGE simulations at $9$ different redshifts we calibrate the semi-analytical HMF model presented in Sec.~\ref{subsec:HMF}; however, there are limitations to this approach that we have to address. First, due to the relatively small box size of $500\,h^{-1}\,\rm Mpc$ the halo catalogs contain only halos with $M \lesssim 4 \times 10^{15}\,h^{-1}\,M_\odot$ depending on cosmology and redshift. Thus the correction function is only calibrated in the mass range that is available from the simulations. 
However, the semi-analytical prediction gives an overall shape of the $f(R)$ HMF. Moreover, the halo mass function is exponentially suppressed at high masses and therefore a correction in this regime has negligible impact on our cosmological analysis. 
The second limitation is the available range for \logfR\ values. FORGE samples \logfR\ in the range $-6.17 < \log_{10}f_{R0} < -4.51$, so we can only predict the parameters of the correction function in this range. Therefore, we adopt a hard upper limit prior of $\logfR = -4.51$ in our analysis. We extend our analysis to $\logfR = -7$ based on the fact that the $f(R)$ HMF approaches the GR HMF for $\logfR \to -\infty$ and thus $a,\ \Delta a \to 0$ and $b \to 1$. Thus, we interpolate the parameters between $\log_{10}f_{R0} = -6.17$ and $\log_{10}f_{R0} = -7$ under the assumption that a model with $\log_{10}f_{R0} = -7$ is indistinguishable from GR with the SPT dataset. Moreover, the correction from the simulation is smaller as we approach GR. We will show in our analysis that the upper bound has no impact on the $f_{R0}$ posterior, as the data strongly disfavor values of $\log_{10}f_{R0} > -4.51$.
Given these limitations on the calibration, we present both the results derived when using the semi-analytical HMF model and when using the FORGE-informed calibrated semi-analytical HMF model.
\newline

In the analysis presented below, we account for the impact of remaining uncertainties in the HMF by following the approach of Ref.~\cite{Costanzi19} and introducing uncertainties in the amplitude and the logarithmic mass trend of the HMF, \ie the HMF with uncertainties given by
    \begin{equation}
        \label{eq:HMF_uncertainty}
         \left. \frac{\dd n}{\dd \mathrm{ln} M} \right\vert_{\mathrm{final}}  =\frac{\dd n}{\dd \mathrm{ln} M} \left(q + s \ln  \left( \frac{M_{200\rm c}}{10^{14}\,h^{-1}M_\odot} \right ) \right) \, ,
    \end{equation}
where $q$ is the uncertainty in the amplitude and $s$ is the uncertainty in the trend with logarithmic mass and we marginalize over the $q$ and $s$ in our analysis (see Sec.~\ref{subsec:likelihood}) 

\section{\label{sec:analysis}Analysis Method}
The method we employ in this work is based on the state-of-the-art weak lensing informed cluster cosmological analysis of the SPT sample \citepalias{Bocquet24Ia}. The method was also used and validated in the recent $f(R)$ gravity forecast \citepalias{Vogt24} for SPT-3G \cite{Benson14} and CMB-S4 \cite{Abazajian19} cluster samples with next-generation weak-lensing data like those expected to come from the Euclid mission \cite{Laureijs11,Scaramella22} or the Vera C.\ Rubin Observatory \cite{Ivezic08,Mandelbaum18}.

\subsection{\label{subsec:obs-mass-rel} Observable--mass relations}
In tSZE cluster surveys, galaxy clusters are identified and selected by observables such as the tSZE detection significance and richness and observable--mass relations link these observables to the halo mass \cite[e.g.,][]{Kaiser86,Angulo12}. Through gravitational weak lensing calibration of these relations, we can relate the observed cluster sample to the HMF, which describes the abundance of halos depending on cosmology, mass, and redshift. 
In this analysis, we employ observable--mass relations that are empirically calibrated with weak-lensing data \cite[e.g.][]{Hoekstra15,Mantz16,Miyatake19,Bellagamba19,Chiu22,Bocquet24II,Grandis24}.
This section outlines the observable--mass relation for tSZE detection significance and optical/NIR richness.

\subsubsection{\label{subsubsec:zeta_mass_rel} tSZE \texorpdfstring{$\zeta$}{zeta}--mass relation}
As in previous SPT studies, we first relate the observed tSZE detection significance \zetahat\ to the intrinsic detection significance $\zeta$ to account for noise in the data. The relation between \zetahat\ and $\zeta$ is given by \citep{Vanderlinde10}
    \begin{equation}
        \label{eq:zeta_to_observed}
        P(\hat\zeta | \zeta) = \mathcal{N} \left( \sqrt{\zeta^2 +3}, 1 \right)  \, .
    \end{equation}
The distribution accounts for the Gaussian noise present in the survey maps, with a correction factor of 3 due to the noise resulting from the matched-filter search for peaks in three dimensions. 
The mean intrinsic tSZE detection significance $\zeta$ is then modeled by
    \begin{equation}
    \begin{split}
        \label{eq:zeta_mass_rel}
        \langle \ln \zeta \rangle = \ln\asz &+ \bsz \ln\left (  \frac{M_{200\mathrm{c}}}{3 \times 10^{14}\, h^{-1} M_\odot} \right) \\
        &+ \csz\ln \left( \frac{E(z)}{E(0.6)} \right) \, ,
    \end{split}
    \end{equation}
where $\asz$, $\bsz$ and $\csz$ are the parameters corresponding to the normalization, mass and redshift trend of the scaling relation and $E(z) = H(z)/H_0$. Adiitionally, we assume a lognormal intrinsic scatter in $\zeta$ with width \sigmalnzeta.
 
Because the SPT surveys vary in depth, and we want to employ one $\zeta$--mass relation for all surveys, the normalization \asz\ and the redshift trend \csz\ are rescaled for each field \citep{Bleem15,Bleem20, Bleem24}, \ie $\zeta_{0,\,\rm field} = \gamma_{\rm field} \asz$ and $\zeta_{z,\,\rm field} = \csz + \rm constant$. In the case of the SPTpol~ECS survey fields the normalization is difficult to calibrate and thus the parameter $\gamma_{\rm ECS}$ is allowed to vary in the analysis \citepalias{Bocquet24Ia}. For the redshift trend \csz\ the variation of the rescaling parameter $c$ across fields is negligible in the SPT-SZ and SPTpol surveys. Therefore, we rescale \csz\ for each survey where the SPT-SZ survey is taken as the Ref.~\cite{Bleem24},
    \begin{equation}
      \begin{split}
        \zeta_\text{z,~SPT-SZ} &= \zeta_\mathrm{z}\, , \\
        \zeta_\text{z,~SPTpol~ECS} &= \zeta_\mathrm{z}-0.09\, , \\
        \zeta_\text{z,~SPTpol~500d} &= \zeta_\mathrm{z}+0.26\, .
      \end{split}
    \end{equation}

\subsubsection{\label{subsubsec:lambda_mass_rel} Cluster richness \texorpdfstring{$\lambda$}{lambda}--mass relation}
As for the tSZE detection significance, the observed richness \lambdahat\ is related to the intrinsic richness $\lambda$ by a Gaussian distribution of the form
        \begin{equation}
        \label{eq:richness_to_observed}
        P(\hat \lambda | \lambda) = \mathcal{N} ( \lambda,  \sqrt{\lambda} )  \, .
    \end{equation}
This relationship accounts for Poisson sampling noise. Note that the Gaussian approximation of a Poisson distribution approximately holds for $\lambda \gtrsim 10$, which is below the richness cut we apply to our sample.
Similar to the $\zeta$--mass scaling relation we assume for the mean intrinsic richness a power law, in mass and $(1+\rm redshift)$:
    \begin{equation}
        \label{eq:lambda_mass_rel}
        \begin{split}
        \langle \ln \lambda \rangle = \ln \alambda &+ \blambda \ln\left (  \frac{M_{200\mathrm{c}}}{3 \times 10^{14}\, h^{-1} M_\odot} \right) \\
        &+ \clambda\ln \left( \frac{1+z}{1.6} \right) \, .
        \end{split}
    \end{equation}
The parameters $\alambda$, $\blambda$ and $\clambda$ govern the normalization, mass and redshift trend, respectively. The cluster intrinsic richness varies around this relation by a log-normal distribution with a width \sigmalnlambda.

As mentioned in Sec.~\ref{subsec:SPT} we use richness measurements from DES for clusters with $z \leq 1.1$ and data from WISE for high-redshift clusters. As matching two distinct types of richness measurements is challenging, we use two separate $\lambda$--mass relations for the DES and WISE data \citepalias{Bocquet24II}, hereafter denoted by subscripts DES and WISE respectively.

\subsection{\label{subsubsec:WL_model} Weak-lensing model in \texorpdfstring{$f(R)$}{f(R)} gravity}
With the above described observable--mass relations, we can model the cluster sample in the $\zeta$--$\lambda$--$z$ space by transforming the halo mass function into the halo observable function and using it to predict cosmological parameters. However, there are no informative priors on the parameters of the $\zeta$ and $\lambda$ scaling relations and their scatters. To empirically calibrate these relations and the corresponding scatters we rely on weak-lensing data.
It has been shown that weak-lensing measurements are a robust way to measure halo masses with well-characterized and controllable biases \citepalias{Bocquet24Ia,Bocquet24II}.

In this analysis, we assume that any $f(R)$-gravity modification of the 
mapping from cluster potential to lensing signal can be neglected, as in previous works \citep{Vogt24,Artis24}. 
First, the $f(R)$ effect on the lensing signal for a given fixed mass distribution is given by a rescaling of the GR signal by a factor of $(1+|f_{R0}|)^{-1}$ \cite{Sotiriou10,Zhang07}, which is negligible for the values of $|f_{R0}|$ we consider in this work. 
Second, while the cluster observables and halo profiles do undergo modifications in $f(R)$ gravity, these effects are small \cite{Schmidt2010,Mitchell18,Ruan24}.
Any changes to the cluster observables $\zetahat$ and $\lambdahat$ will be accounted by the empirical calibration of the observable mass relations. Changes to the halo profiles are more concerning, because they could impact the weak lensing inferred cluster masses. Accounting for these effects self-consistently within the weak lensing model described below would require a study of the halo shapes using $f(R)$ numerical simulations and measurement of any changes in the inferred weak-lensing masses. Because we know these effects are smaller than the current uncertainties on the weak lensing model, which are dominated by uncertainties in the hydrodynamical effects and on photometric redshift systematics, we adopt the GR-based calibration of the weak lensing model presented in \citetalias{Bocquet24Ia} and described below.

\subsubsection{\label{subsubsec:DES_WL} DES weak-lensing model}
The model we adopt for DES weak-lensing data was studied and described in detail in \citetalias{Bocquet24Ia} and works referenced therein. Here we provide a summary of the method. The weak-lensing observable is the reduced tangential shear profile, which is related to the underlying projected halo mass distribution $\Sigma$ by
    \begin{equation}
      \label{eq:g_model}
      g_\mathrm{t}(r,\MWL) = \frac{\Delta\Sigma(r,\MWL)~\Sigma_\mathrm{crit}^{-1}}{1-\Sigma(r,\MWL)~\Sigma_\mathrm{crit}^{-1}} \, .
    \end{equation}
Here $\Delta\Sigma(r)\equiv\langle\Sigma(<r)\rangle-\Sigma(r)$ is the surface density contrast and $\Sigma_\mathrm{crit}^{-1}$ is the lensing efficiency or inverse critical surface mass density, given by
    \begin{equation} \label{eq:Sigmacrit}
        \Sigma_\mathrm{crit}^{-1}= \frac{4\pi G}{c^2} \frac{D_\mathrm{l}}{D_\mathrm{s}} \times
        \mathrm{max}\left[0, D_\mathrm{ls} \right] \, ,
    \end{equation}
where $c$ is the speed of light and $D_{\rm s}$, $D_{\rm l}$, $D_{\rm ls}$ are the angular diameter distances between the observer and the source, the observer and the lens, and the source and the lens, respectively. We model $\Sigma$ by the line of sight integral of a Navarro-Frenk-White profile (NFW) \cite{Navarr01996,Bartelmann17} and we refer to the associated mass as the weak-lensing mass, \MWL. We account for possible miscentering of the selected cluster center by assuming a constant density within the cluster miscentering radius $R_{\rm min}$, \ie $\Sigma(R) = \Sigma(R_{\rm min})$ for $R \leq R_{\rm min}$ (see \citetalias{Bocquet24Ia}~Sec.~IVC for more details). Cluster member contamination $f_\mathrm{cl}(r)$ is corrected by a factor $(1-f_\mathrm{cl}(r))$ to the reduced tangential shear profile \citepalias{Bocquet24Ia}. 

Given that the cluster profiles are not perfectly described by an NFW profile, the computed weak-lensing mass \MWL\ is a biased and noisy estimator of the true cluster mass $M_{200 \rm c}$ \citep{becker&kravtsov11,oguri&hamana11}. To account for the bias we use a scaling relation between \MWL\ and $M_{200 \rm c}$ with a mean relation of \cite{Grandis21}
    \begin{small}
    \begin{equation}
        \label{eq:WL_mass_rel}
         \left\langle \mathrm{ln} \left( \frac{M_{\mathrm{WL}}}{M_0}  \right)  \right\rangle =\bWL(z) + \bWLM \mathrm{ln} \left( \frac{M_{200\mathrm{c}}}{M_0} \right) \,.
    \end{equation}
    \end{small}
Here \bWL\ is the logarithmic mass bias normalization and \bWLM\ is the mass trend in this bias at a pivot mass $M_0 = 2 \times 10^{14}\, h^{-1} M_\odot$. We assume a log-normal scatter of the true relations with a width described by
    \begin{equation}
        \label{eq:WL_mass_var}
       \mathrm{ln}\,\sWLall^2 =
       \sWL(z)  +\sWLM  \mathrm{ln}  \left( \frac{M_{200\mathrm{c}}}{M_0} \right)\, ,
    \end{equation}
where \sWL\ is the normalization and \sWLM\ is the mass trend of the scatter.

The parameters of the above mean scaling relation and scatter are calibrated from simulations by extracting the weak-lensing inferred mass from hydrodynamical simulations and calculating the corresponding cluster mass from the matched $N$-body simulation at different redshifts \cite{Grandis21}. 
The calibration results in a mean value and uncertainty obtained from the posterior for each of the above parameters.

In this model, the logarithmic mass bias normalization, \bWL, and the normalization of the scatter, \sWL, are functions of redshift and calibrated from the simulations at four redshift values: $z\in\{0.252, 0.470, 0.783, 0.963\}$. Therefore, we model the two parameters in the analysis as
    \begin{equation}
        \label{eq:WL_param_modeling}
        p = \mathcal{N}(\bar p, (\Delta p)^2) = \bar p(z) + \Delta p(z)\, \mathcal{N}(0, 1)\, ,
    \end{equation}
where $\bar p(z)$ is the mean value and $\Delta p(z)$ is the uncertainty of the corresponding parameter $p(z)$ at redshift $z$. We interpolate linearly to obtain the values for these parameters at any intermediate redshift. To accurately describe the uncertainty of the logarithmic mass bias $\bWL(z)$, the uncertainty in this parameter, $\Delta \bWL(z)$, is modeled as a linear combination of two redshift-dependent components, which are both interpolated in the considered redshift range based on the values in Tab.~\ref{tab:WLmodel}
\citepalias{Bocquet24Ia}:
    \begin{equation}
        \Delta \bWL(z) =  \Delta_1 \bWL(z) +  \Delta_2 \bWL(z).
    \end{equation}
The values of the bias and scatter normalization parameters, Eqs.~\eqref{eq:WL_mass_rel} and \eqref{eq:WL_mass_var}, as well as their uncertainties at the simulation redshifts used in this work are summarized in Table~\ref{tab:WLmodel}. 
The uncertainties of these parameters include various elements such as uncertainties from baryonic effects, photo-$z$ calibration, miscentering and shear calibration \cite{Grandis21}. The total uncertainty is primarily influenced by uncertainties in baryonic effects at low redshifts, while at high redshifts, the uncertainty in photo-$z$ calibration becomes dominant. Overall uncertainty from the weak-lensing model remains small across the calibrated redshift range, contributing to approximately $1\,\%$ of the total uncertainty (see Fig.~10 in \citetalias{Bocquet24Ia}).
\begin{table}
    \caption{\label{tab:WLmodel}
    Normalization and uncertainties ($\Delta$) of the amplitude and scatter of the weak-lensing-mass-to-halo-mass relation derived from the simulations at redshifts $z\in\{0.252, 0.470, 0.783, 0.963\}$.}
    \begin{ruledtabular}
    \begin{tabular}{l S[table-format=-1.3] S[table-format=-1.3] S[table-format=-1.3] S[table-format=-1.3]}
    Parameter & \multicolumn{1}{c}{$z_0$} & \multicolumn{1}{c}{$z_1$} & \multicolumn{1}{c}{$z_2$} & \multicolumn{1}{c}{$z_3$} \\
    \colrule
    $\bWL(z)$          & -0.042  & -0.040  & -0.033  & -0.082  \\
    $\Delta_1 \bWL(z)$ & -0.006  & -0.014  & -0.052  & -0.112  \\
    $\Delta_2 \bWL(z)$ &  0.008  &  0.015  &  0.017  & -0.010  \\
    $\sWL(z)$          & -3.115  & -3.074  & -2.846  & -1.945  \\
    $\Delta \sWL(z)$   &  0.044  &  0.048  &  0.060  &  0.101  \\
    \end{tabular}
    \end{ruledtabular}
\end{table}

\subsubsection{\label{subsubsec:HST_WL} HST weak-lensing model}
A similar model is applied to the HST-39 dataset.
The shear profiles from HST are modeled by the line of sight integral of an NFW profile with a concentration from Ref.~\cite{diemer&joyce19}. From the NFW a weak-lensing mass $\MWL$ is calculated and related to the true halo mass with a mean relation \cite{Schrabback18},
    \begin{equation}
        \label{eq:HST_WL}
        \langle \ln\,\MWL \rangle = \bWLHST + \ln\,M_{200\mathrm{c}}\, .
    \end{equation}
The true relation scatters around the mean by a Gaussian distribution with width \sWLHST.
The scatter \sWLHST\ accounts for all sources of uncertainties in the \MWL--$M_{200\rm c}$ relation. Here each cluster has its own bias and scatter and associated uncertainties by calibrating Eq.~\eqref{eq:HST_WL} for each cluster individually. We refer the reader to the original works for a more detailed explanation of the cluster lensing model employed in the HST dataset \cite{Schrabback18,Schrabback21,Zohren22,Sommer22}.

\subsection{\label{subsubsec:multi_obs} Multivariate observable--mass relation}
To account for possible correlation among the three observables, unbiased tSZE detection significance $\zeta$, intrinsic richness $\lambda$ and weak-lensing mass \MWL\, we employ the multivariant observable--mass relation from the work of \citetalias{Bocquet24Ia}. For this, the lognormal scatters of the observables, \sigmalnzeta, \sigmalnlambda\ and \sWLall\ are combined into a covariance matrix of the form
\begin{equation} 
  \label{eq:covmat}
  \begin{footnotesize}
        \boldsymbol{\Sigma} = 
        \begin{pmatrix}
          \sigma_{\ln\zeta}^2 & \rho_\mathrm{SZ,WL}\sigma_{\ln\zeta}\sWLall & \rho_\mathrm{SZ,\lambda}\sigma_{\ln\zeta}\sigma_{\ln\lambda} \\
          \rho_\mathrm{SZ,WL}\sigma_{\ln\zeta}\sWLall & \sWLall^2 & \rho_\mathrm{WL,\lambda}\sWLall\sigma_{\ln\lambda} \\
\rho_\mathrm{SZ,\lambda}\sigma_{\ln\zeta}\sigma_{\ln\lambda} & \rho_\mathrm{WL,\lambda}\sWLall\sigma_{\ln\lambda} & \sigma_{\ln\lambda}^2
        \end{pmatrix},
  \end{footnotesize}
\end{equation}
where $\rho_\mathrm{SZ,\lambda},\  \rho_\mathrm{SZ,WL}$ and $\rho_\mathrm{WL,\lambda}$ are the correlation coefficients between $\zeta$ and \MWL, $\zeta$ and $\lambda$, and $\lambda$ and \MWL\ respectively.
The joint multi-observable--mass relation is then given by a multivariate Gaussian with correlation matrix $\boldsymbol{\Sigma}$ 
    \begin{equation}
    \begin{split}
      \label{eq:scaling_relation}
       & P\Bigl(
        \begin{bmatrix}
          \ln\zeta \\ \ln \MWL \\ \ln\lambda
        \end{bmatrix} \big| M,z,\boldsymbol p\Bigr) = \\
        & \hspace{3cm} \mathcal N\Bigl(
        \begin{bmatrix}
          \langle\ln\zeta\rangle(M,z,\boldsymbol p) \\ \langle\ln M_\mathrm{WL}\rangle(M,z,\boldsymbol p) \\ \langle \ln\lambda\rangle(M,z,\boldsymbol p)
        \end{bmatrix}, \boldsymbol{\Sigma}\Bigr)\, .
    \end{split}
    \end{equation}
%

\subsection{\label{subsec:likelihood} Likelihood and priors}
The analysis relies on Bayesian statistics and we obtain cosmological and scaling relation parameters $\boldsymbol{p}$ using a cluster population model. The likelihood model employed in this analysis is based on the recent $\Lambda$CDM SPT$\times$DES+HST analysis of \citetalias{Bocquet24Ia,Bocquet24II} and was verified for an $f(R)$ cosmology in our forecast work \citepalias{Vogt24}. Following \citetalias{Bocquet24Ia} and \citetalias{Bocquet24II}, we approximate the multi-observable cluster abundance likelihood with a Poissonian distribution:
    \begin{equation}
    \begin{split}
      \label{eq:poisson_likelihood}
        \ln \mathcal L(\boldsymbol p) &= \sum_i \ln \int_{\hat\lambda_\mathrm{min}}^\infty \dd\hat\lambda \frac{\dd^3 N(\boldsymbol p)}{\dd\hat\zeta \, \dd\hat\lambda \, \dd z} \Big|_{\hat\zeta_i, z_i} \\
        &- \int_{z_\mathrm{min}}^{z_\mathrm{max}} \dd z \int_{\hat\zeta_\mathrm{min}}^\infty \dd\hat\zeta \int_{\hat\lambda_\mathrm{min}}^\infty \dd\hat\lambda \frac{\dd^3 N(\boldsymbol p)}{\dd\hat\zeta \, \dd\hat\lambda \, \dd z } \\
       &+ \sum_i \ln\frac{\frac{\dd^4 N(\boldsymbol p)}{\dd\hat\zeta \, \dd\hat\lambda \, \dd \boldsymbol g_\mathrm{t} \dd z}
        \Big|_{\hat\zeta_i, \hat\lambda_i, \boldsymbol{g}_{\mathrm{t},i}, z_i}}
        {\int_{\hat\lambda_\mathrm{min}}^\infty \dd\hat\lambda \frac{\dd^3 N(\boldsymbol p)}{\dd\hat\zeta \, \dd\hat\lambda \, \dd z}
        \Big|_{\hat\zeta_i, z_i}}
        + \mathrm{const.} \, ,
    \end{split}
    \end{equation}
where both sums run over all clusters $i$. The differential cluster numbers $\frac{\dd^3N}{ \dd\,\mathrm{obs}}$ in the above likelihood are the differential halo observable function (HOF) and is given in the $\hat\zeta$--$\hat\lambda$--$z$ space by
    \begin{equation}
        \label{eq:HOF_with_3_obs}
        \begin{split}
        \frac{\dd^3 N (\boldsymbol p)}{\dd\hat\zeta \, \dd\hat\lambda \, \dd z } &=
        \int \dd\Omega_\mathrm{s} 
        \iiint  \dd M\, \dd\lambda\, \dd\zeta\, P(\hat\zeta|\zeta) P(\hat\lambda|\lambda)\\
        &P(\zeta, \lambda |M,z,\boldsymbol p) 
          \frac{\dd^2 N (M, z, \boldsymbol p)}{\dd M \, \dd V} 
         \frac{\dd^2 V(z,\boldsymbol p)}{\dd z \, \dd\Omega_\mathrm{s}} 
         \, ,
        \end{split}
    \end{equation}
and in the $\zetahat$--$\lambdahat$--$\boldsymbol g_\mathrm{t}$--$z$ space by
%
    \begin{equation} 
        \label{eq:HOF_with_4_obs}
        \begin{split}
        \frac{\dd^4 N(\boldsymbol p)}{\dd\hat\zeta \, \dd\hat\lambda \, \dd \boldsymbol g_\mathrm{t} \dd z }& = 
        \int \dd\Omega_\mathrm{s}
        \iiiint  \dd M\, \dd\zeta\, \dd\lambda\, \dd M_\mathrm{WL} \\
        &P(\boldsymbol g_\mathrm{t}|M_\mathrm{WL}, \boldsymbol p)
        P(\hat\zeta|\zeta) P(\hat\lambda|\lambda) \\
       & P(\zeta, \lambda, M_\mathrm{WL} |M,z,\boldsymbol p) \\
        &  \frac{\dd^2 N (M, z, \boldsymbol p)}{\dd M \, \dd V} 
         \frac{\dd^2 V(z,\boldsymbol p)}{\dd z \, \dd\Omega_\mathrm{s}} 
        \, .
        \end{split}        
    \end{equation}
%
Here $\Omega_\mathrm{s}$ is the survey solid angle, the factors $\frac{\dd^2 N (M, z, \boldsymbol p)}{\dd M \dd z}$ and $\frac{\dd^2 V (z, \boldsymbol p)}{\dd z \dd\Omega_\mathrm{s}}$ are the HMF and the differential volume element for the corresponding cosmology whereas $P(\hat\zeta|\zeta)$ and $P(\hat\lambda|\lambda)$ relate the observed quantity to the intrinsic one given by Eqs.~\eqref{eq:zeta_to_observed} and \eqref{eq:richness_to_observed} respectively. Moreover, $P(\zeta, \lambda |M,z,\boldsymbol p)$ and $P(\zeta, \lambda, M_\mathrm{WL} |M,z,\boldsymbol p)$ are obtained from the multivariant observable--mass relation Eq.~\eqref{eq:scaling_relation}, and $P(\boldsymbol g_\mathrm{t}|M_\mathrm{WL}, \boldsymbol p)$ is given by the product of Gaussian probabilities in each radial bin $i$ of the tangential shear profiles $g_\mathrm{t}$ with the shape noise $\Delta g_{\mathrm{t},i}$
%
    \begin{equation}
    \begin{split}
        \label{eq:lensing_likelihood}
        P(\boldsymbol g_\mathrm{t}|M_\mathrm{WL}, \boldsymbol p) &= \prod_i \left(\sqrt{2\pi}\Delta g_{\mathrm{t},i} \right)^{-1} \\
        &\exp \left[ -\frac12 \left(\frac{g_{\mathrm{t},i} - g_{\mathrm{t},i}(M_\mathrm{WL}, \boldsymbol p)}{\Delta g_{\mathrm{t},i}}\right)^2 \right] \, .
    \end{split}
    \end{equation}
%
Note that the first two terms of the total Poisson likelihood, Eq.~\eqref{eq:poisson_likelihood}, are independent of the weak-lensing data and thus are associated with the Poisson likelihood of the cluster sample in the $\zetahat$--$z$ space with the condition $\lambdahat > \lambdahat_{\rm min}(z)$. The last term in Eq.~\eqref{eq:poisson_likelihood} includes the information of the mass calibration from the weak-lensing data and is therefore often called the mass-calibration likelihood.
The likelihood is implemented in the \textsc{CosmoSIS} framework as a Python module \citep{Zuntz15} and we use the nested sampling algorithm \textsc{nautilus} to run the MCMC chains \citep{Lange23}. 
 
Given the large number of cosmological and nuisance parameters considered in the analysis, we combine the SPT cluster dataset with the primary CMB data from Planck (Planck\,2018 TT,TE,EE+lowE) \cite{Planck2020} to break parameter degeneracies and recover meaningful constraints on $f(R)$ gravity.
This combination of data is sound, because the standard cosmological analysis from \citetalias{Bocquet24II} showed no statistically significant tension between the SPT-clusters$\times$WL dataset and the Planck data.

We emphasize that primary CMB data like those from Planck\,2018 place only weak constraints on $f(R)$ gravity of the order of $\log_{10}|f_{R0}| \lesssim -3$ \cite{Planck15,Kou23}, and thus the constraints in \logfR\ are primarily coming from the SPT cluster sample.
However, CMB data are essential to constrain the remaining cosmological parameters such as $\Omega_{\rm m}h^2$.
The Planck\,2018 likelihood is implemented in \textsc{CosmoSIS}, and because one can assume that the cluster likelihood and the Planck\,2018 likelihood are independent, we multiply the two likelihoods in a joint analysis. We account for the effect of $f(R)$ gravity in the Planck\,2018 likelihood by using the $f(R)$ power spectrum computed with \textsc{MGCAMB} \cite{Zhao09,Hojjati11,Zucca19,Wang23}. 
    \begin{table}
      \caption{Parameters and priors of our $f(R)$ gravity analysis of the DES and HST weak-lensing informed SPT cluster sample and Planck\,2018. 
      The prior on $\Omega_\nu h^2$ corresponds to a prior on the sum of neutrino masses $\Sigma m_\nu \sim \mathcal U(0, 0.6)~\mathrm{eV}$.}
      \label{tab:priors}
      \begin{ruledtabular}
        \begin{tabular}{lll}
          Parameter & Description & Prior\\
          \colrule
          \multicolumn{3}{l}{DES Y3 cluster lensing} \\
          $\Delta_1\bWL$ & scaling of bias & see Table~\ref{tab:WLmodel} \\
          $\Delta_2\bWL$ & scaling of bias & see Table~\ref{tab:WLmodel} \\
          $\bWLM$ & mass slope of bias & $\mathcal N(1.029, 0.006^2)$ \\
          $\Delta \sWL$ & scaling of scatter & see Table~\ref{tab:WLmodel} \\
          $\sWLM$ & mass slope of scatter & $\mathcal N(-0.226, 0.040^2)$ \\
          \colrule
          \multicolumn{3}{l}{HST cluster lensing} \\
          $\Delta \bWLHST$ & amplitude of bias & varied by cluster \\
          $\Delta \sWLHST$ & amplitude of scatter & varied by cluster\\
          \colrule
          \multicolumn{2}{l}{tSZE--mass parameters} \\
          $\ln$\asz & amplitude & $\mathcal{U} (0.39, 0.93)$ \\
          \bsz & mass slope & $\mathcal{U} (1.56, 1.9)$\\
          \csz & redshift evolution & $\mathcal{U} (0.1, 1.25)$\\
          \sigmalnzeta & intrinsic scatter & $\mathcal{U} (0.003, 0.4)$\\
          $\gamma_\mathrm{ECS}$ & depth of SPTpol ECS & $\mathcal{U} (0.9, 1.2)$\\
          \colrule
          \multicolumn{3}{l}{DES richness--mass parameters (used for $z<1.1$)} \\
          $\ln \lambda_{0{,\,\rm DES}}$ & amplitude & $\mathcal{U} (3.5, 3.9)$\\
          $\lambda_{M,\,\rm DES}$ & mass slope & $\mathcal{U} (1.08, 1.42)$\\
          $\lambda_{z,\,\rm DES}$ & redshift evolution &$\mathcal{U} (-0.34, 0.8)$\\
          $\sigma_{\rm ln \lambda,\,\rm DES}$ & intrinsic scatter & $\mathcal{U} (0.01, 0.33)$\\
          \colrule
          \multicolumn{3}{l}{WISE richness--mass parameters (used for $z>1.1$)} \\
          $\ln \lambda_{0,\,\rm WISE}$  & amplitude & $\mathcal{U} (3.48, 5.12)$\\
          $\lambda_{M,\,\rm WISE}$  & mass slope & $\mathcal{U} (0.6, 1.33)$\\
          $\lambda_{z,\,\rm WISE}$  & redshift evolution & $\mathcal{U} (-4.27)$\\
          $\sigma_{\rm ln \lambda,\,\rm WISE}$ & intrinsic scatter & $\mathcal{U} (0.005, 0.34)$\\
          \colrule
          \multicolumn{3}{l}{Correlation coefficients} \\
          $\rho_\mathrm{SZ,WL}$ & tSZE--weak-lensing & $\mathcal U(-0.5, 0.5)$ \\
          $\rho_\mathrm{SZ,\lambda}$ & tSZE--richness & $\mathcal U(-0.5, 0.5)$ \\
          $\rho_\mathrm{WL,\lambda}$ & weak-lensing--richness & $\mathcal U(-0.5, 0.5)$ \\
          \colrule
          \multicolumn{3}{l}{HMF uncertainty} \\
          $q$ & bias uncertainty & $\mathcal{N} (1, 0.2^2)$ \\
          $s$ & slope uncertainty & $\mathcal{N} (0, 0.1^2)$ \\
          \colrule
          \multicolumn{3}{l}{Cosmology} \\
          $\Omega_{\rm m}$ & matter density & $\mathcal{U}(0.27, 0.36)$ \\
          $\Omega_\nu h^2$ & neutrino density & $\mathcal U(0, 0.00644)$ \\
          $\Omega_{\rm b}h^2$ & baryon density & $\mathcal U (0.02191,0.02281)$ \\
          $h$ & Hubble parameter & $\mathcal U(0.643, 0.702)$ \\
          $\ln10^{10}A_s$ & amplitude of $P(k)$ & $\mathcal{U} (2.89, 3.1)$ \\
          $n_s$ & scalar spectral index & $\mathcal U(0.9517, 0.9781)$ \\
          $\tau$ & depth of reionization & $\mathcal U(0.02, 0.08)$\\
          \logfR & $f(R)$ gravity parameter & $\mathcal U(-7, -3)$ 
        \end{tabular}
      \end{ruledtabular}
    \end{table}

In this analysis, we vary 23 nuisance parameters and eight cosmological parameters.
All parameters with their priors are listed in Table~\ref{tab:priors}.
For the standard cosmological parameters, we adopt uniform priors with ranges that are based on the Planck\,2018 posteriors, since these parameters will be best constrained by the Planck\,2018 dataset. 
For the $f(R)$ gravity parameter \logfR\ we apply a uniform prior $\mathcal{U}(-7, -3)$. 
Note that the GR limit at $|f_{R0}| = 0$ cannot be reached when using a logarithmic prior and thus would introduce an infinitely large parameter volume below our lower bound when we calculate the upper limit of the $f(R)$ parameter. To avoid the sensitivity of the upper limit of the $f(R)$ parameter on the choice of the lower prior boundary, we transform the parameter space of the final output chain from logarithmic to linear. In linear space, the parameter volume between 0 and $10^{-7}$ is negligible. 
We account for the uncertainty in the weak-lensing model as described in Sec.~\ref{subsubsec:DES_WL} by Gaussian priors on these parameters.
To ensure a positive-definite covariance matrix of the multivariate observable--mass relation we assume priors on the correlation coefficients of $\mathcal{U}(-0.5, 0.5)$. No informative priors are applied for the tSZE and richness observable--mass relation parameters, and we adopt sufficiently wide uniform priors for these parameters. To account for systematic uncertainties in the HMF, we use Gaussian priors on the amplitude and slope of the HMF as defined in Eq.~\eqref{eq:HMF_uncertainty}.

\section{\label{sec:results}Results}
In this section, we present our constraints on the $f(R)$ gravity model derived from the analysis of DES and HST weak lensing calibrated SPT clusters combined with Planck\,2018 data. For our baseline analysis, we employ the FORGE-calibrated HMF model, because it is considered to be more accurate due to its empirical calibration with the FORGE numerical simulations. For comparison, we also include results obtained using the uncorrected semi-analytical HMF model. 
Note that all reported upper limits are provided at the $95\,\%$ credible level and uncertainties at $68\,\%$ credibility. 

\subsection{\label{subsec:results_comp_LCDM}Comparison to \texorpdfstring{$\Lambda$}{L}CDM results}
\begin{figure*}
        \centering
        \includegraphics[width=\textwidth]{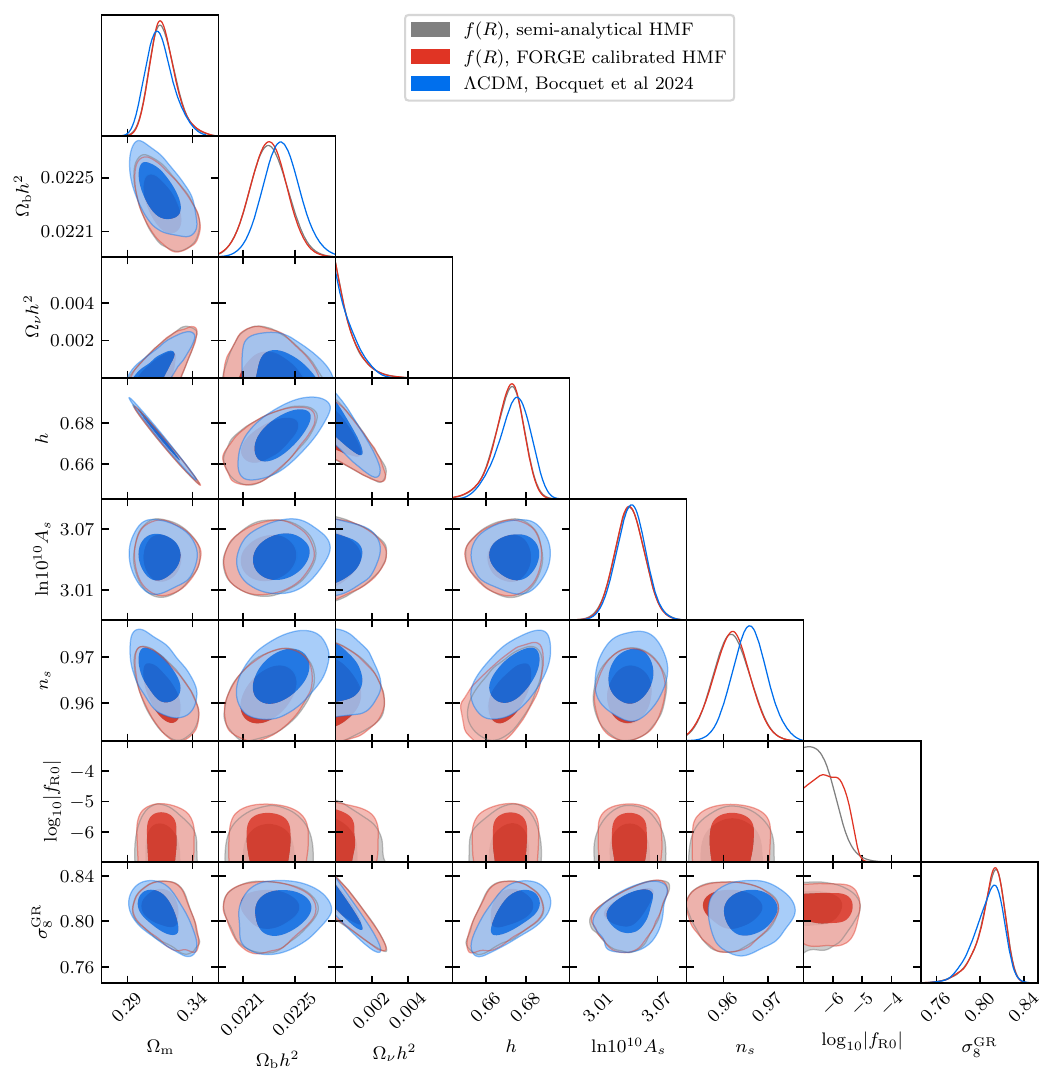}
        \vskip-0.25cm
        \caption{Posterior distribution of the cosmological parameters for our $f(R)$ analysis (gray and red) and the $\Lambda$CDM analysis from \citetalias{Bocquet24II} (blue) of the DES and HST weak-lensing mass calibrated SPT clusters combined with Planck\,2018 CMB data.
        Because the two $f(R)$ HMF models give similar results, the gray contours from the semi-analytical HMF (see Sec.~\ref{subsec:HMF}) are hidden by the red ones from the FORGE calibrated HMF (see Sec.~\ref{subsec:HMF_cali}). As expected, the results from the analysis show consistent results in the non-$f(R)$ cosmological parameters.}
        \label{fig:comp_results}
    \end{figure*}

We first compare our results to those from the $\Lambda$CDM analysis of \citetalias{Bocquet24II}. Because we report a tight upper limit on \logfR\ (see next section) and thus the deviation from a $\Lambda$CDM cosmology is relatively small, we expect consistent results between the two analyses for all standard cosmological parameters.
Figure~\ref{fig:comp_results} shows the posterior distribution of the cosmological parameters from both analyses. Overall, the standard cosmological parameters are in good agreement with the $\Lambda$CDM results. The only discernible parameter shifts are observed in $\Omega_{\rm b}h^2$ and $n_s$, which are both primarily constrained by the Planck\,2018 data. However, the shift of the two parameters is within the $68\,\%$ credible contours and within the statistical uncertainties. An explanation for the shift can be given by the fact that $\Omega_{\rm b}h^2$ and $n_s$ as well as \logfR\ change the power spectrum on small scales, and in opposite directions. 

A complete comparison of all parameters from the $f(R)$ and $\Lambda$CDM analysis can be found in Appendix~\ref{app:comp_results}, Fig.~\ref{fig:comp_all}. We summarize in Table~\ref{tab:comp_LCDM} the constraints on the $\zeta$--mass and $\lambda$--mass relation parameters, as well as on the correlation coefficients for the $f(R)$ and $\Lambda$CDM analyses. The results are in good agreement with $\Lambda$CDM, indicating no significant deviations in the observable--mass relations under the $f(R)$ gravity model. 
Compared to the $\Lambda$CDM constraints, we obtain slightly weaker constraints for most of the observable--mass relation parameters (see the third column of Table~\ref{tab:comp_LCDM}). 
The largest increase in the uncertainties is observed in the $\zeta$--mass relation parameters, the amplitude mass trend of the $\lambda$--mass relation for the DES richness, and the scatter of the $\lambda$--mass relation for the WISE richness. 
An overall increase in the uncertainties is expected due to the additional degrees of freedom introduced by the $f(R)$ gravity model.

We observe a $1.5\,\sigma$ shift in the correlation parameter between the tSZE detection significance and richness, $\rho_{SZ,\lambda}$, compared to the $\Lambda$CDM results, which is attributed to different models for the scatter in richness assumed in the two analyses:
in the $\Lambda$CDM analysis, a lognormal richness scatter with width $\lambda^{-1/2}$ was used, which resulted in a negative correlation between the tSZE detection significance and richness. 
In this $f(R)$ analysis, we use a Gaussian approximation of Poisson noise (see Eq~\eqref{eq:richness_to_observed}) and report a vanishing correlation 
between the two quantities. 
To check if this is due to the different scatter models, we also run the analysis with the richness scatter model as in the $\Lambda$CDM analysis, and we obtained the same result as the $\Lambda$CDM analysis for the correlation parameter $\rho_{SZ,\lambda}$.
The same is true for the scatter parameter of the WISE richness $\sigma_{\rm ln \lambda,\,\rm WISE}$.
    \begin{table}
      \caption{Constraints on the observable--mass relation parameters, the scatter correlation coefficients and cosmological parameters for the $f(R)$ and $\Lambda$CDM analyses in the second and third column respectively. The last column shows the relative increase of the parameter uncertainties in the $f(R)$ analysis with respect to $\Lambda$CDM. Note that all reported upper limits are provided at the $95\,\%$ credible level and uncertainties at $68\,\%$ credibility. A missing entry (\dots) for the increase in the uncertainty indicates no change in the error of this parameter.}
      \label{tab:comp_LCDM}
      \begin{ruledtabular}
        \begin{tabular}{lccc}
          Parameter & $f(R)$ & $\Lambda$CDM &increase in the\\
          &&&uncertainties\\
          \colrule
          \multicolumn{4}{l}{tSZE--mass parameters} \\
          $\ln$\asz & $0.69\pm 0.09$ & $0.69\pm0.06$ & $50\,\%$\\ 
          \bsz & $1.73\pm 0.05$ & $1.73\pm0.04$ & $25\,\%$\\ 
          \csz & $0.73\pm 0.13$   & $0.74\pm0.11$ & $18\,\%$\\ 
          \sigmalnzeta & $0.22\pm 0.06$ & $0.21\pm0.05$ & $20\,\%$\\ 
          $\gamma_\mathrm{ECS}$ & $1.05\pm 0.03$ & $1.05\pm0.03$ & \dots \\ 
          \colrule
          \multicolumn{4}{l}{DES richness--mass parameters (used for $z<1.1$)} \\
          $\ln \lambda_{0,\,\rm DES}$ & $3.73\pm 0.06$ & $3.73\pm0.05$ & $20\,\%$\\ 
          $\lambda_{M,\,\rm DES}$ & $1.23\pm 0.05$ & $1.25\pm0.04$ & $25\,\%$\\ 
          $\lambda_{z,\,\rm DES}$ & $0.13\pm 0.13  $ & $0.15\pm0.12$ & $9\,\%$\\ 
          $\sigma_{\rm ln \lambda,\,\rm DES}$ & $0.21\pm 0.04$ & $0.18\pm0.04$ & \dots  \\ 
          \colrule
          \multicolumn{4}{l}{WISE richness--mass parameters (used for $z>1.1$)} \\
          $\ln \lambda_{0,\,\rm WISE}$  & $4.30\pm 0.21$ & $4.33\pm0.21$ & \dots  \\ 
          $\lambda_{M,\,\rm WISE}$  &  $1.0\pm0.1$& $0.96\pm0.09$ & $10\,\%$\\ 
          $\lambda_{z,\,\rm WISE}$  & $-2.0\pm 0.6$ & $-2.0\pm0.6$ & \dots\\ 
          $\sigma_{\rm ln \lambda,\,\rm WISE}$ & $0.14\pm 0.07$ & $0.12\pm0.05$ & $40\,\%$\\ 
          \colrule
          \multicolumn{4}{l}{Correlation coefficients} \\
          $\rho_\mathrm{SZ,WL}$ & $< 0.22 $ & $<0.17$ & $30\,\%$\\ 
          $\rho_\mathrm{SZ,\lambda}$ & $0.03\pm0.35$ & $<0.08$& \dots \\ 
          $\rho_\mathrm{WL,\lambda}$ & $-0.05\pm0.31$ & $-0.10\pm0.24$  & $30\,\%$\\ 
          \colrule
          \multicolumn{4}{l}{Cosmology} \\
          $\Omega_{\rm m}$ & $0.318\pm0.011$ & $0.315\pm 0.011$ & \dots \\
          $\Omega_\nu h^2$ & $<0.2$  & $<0.18$ & $10\,\%$\\
          $\Omega_{\rm b}h^2$ & $0.0223\pm 0.0002$ & $0.0224\pm 0.0002$  & \dots \\
          $h$ & $0.671\pm0.009$ & $0.674\pm0.008$ & $13\,\%$\\
          $\ln10^{10}A_s$ & $3.042\pm0.015$ & $3.043 \pm 0.015$ & \dots \\
          $n_s$ & $0.962 \pm 0.004$ & $0.966 \pm 0.004$ & \dots \\
          \logfR & $< -5.32$ & \dots & \dots \\
          $\sigma_8^{\rm GR}$ & $0.809\pm0.015$ & $0.807\pm0.013$ & $16\,\%$
        \end{tabular}
      \end{ruledtabular}
    \end{table}

\subsection{\label{subsec:results_fRM}\texorpdfstring{$f(R)$}{f(R)} gravity constraints}
Our $f(R)$ analysis of the combination of weak-lensing mass calibrated SPT clusters and the Planck\,2018 dataset results in the current tightest constraints available from clusters and CMB data and other probes of the large-scale structure. With our baseline analysis using the FORGE-calibrated HMF, we obtain an upper bound on the $f(R)$ parameter:
    \begin{equation}
        \label{fR0_constraints_FORGE}
        \logfR < -5.32 \quad (95\,\%\ \mathrm{credible\ level})\, .
    \end{equation}
This result is consistent with a $\Lambda$CDM cosmology and excludes all $f(R)$ parameter space that does not lead to substantial screening of halos. When applying the semi-analytical HMF model, we achieve a $3\,\%$ tighter upper bound on \logfR:
    \begin{equation}
        \label{fR0_constraints_semi}
        \logfR < -5.46 \quad (95\,\%\ \mathrm{credible\ level})\, .
    \end{equation}
The slightly stronger constraint from the semi-analytical HMF model can be explained by the larger enhancement in the halo mass function compared to the FORGE simulations, see Sec.~\ref{subsec:HMF_cali}. Therefore, one expects a greater sensitivity to $\logfR$, which leads to tighter constraints on the $f(R)$ parameter. The remaining cosmological parameter posteriors are found to be consistent with those in the $\Lambda$CDM analysis. This is because we use Planck\,2018 data in our analysis, which tightly constrains the $\Lambda$CDM cosmological parameters and thus eliminates potential degeneracies. Moreover, degeneracies between \logfR\ and other cosmological parameters that have previously been found are more pronounced for higher values of \logfR, which are excluded by our dataset \cite{Harnois23,Baldi14,Hagstotz19}.

For comparison, the best previous constraints on $f(R)$ gravity from clusters are presented in Ref.~\cite{Cataneo14}. The authors obtained $\logfR < -4.79$ using clusters from ROSAT and the Massive Cluster Survey, combined with primary CMB, SN, and BAO data. Our analysis improves upon this result by a factor of 3.4 in $f_{R0}$ without using any information from SNe or BAO. The improved constraints are due to the large cluster sample and the weak lensing mass calibration dataset that we use in this analysis, with $1,005$ clusters compared to 224 clusters in the analysis of Ref.~\cite{Cataneo14}.
Recent results from the eROSITA cluster analysis reported an upper limit of $\logfR < -4.12$, considering the neutrino mass as a free parameter \cite{Artis24}. Although this constraint is significantly weaker than ours, it is important to note that their analysis was based on clusters alone. eROSITA can place meaningful constraints from the clusters alone, because of the larger weak-lensing calibrated cluster sample of $5,259$ clusters and lower redshift range for which the effect of $f(R)$ gravity is larger. On the other hand, they employed a different HMF based on the model of Ref.~\cite{Hagstotz19}, which predicts a slightly smaller enhancement of the $f(R)$ HMF. For a fairer comparison with the eROSITA results, we also perform our analysis using the HMF of Ref.~\cite{Hagstotz19} but still including Planck\,2018 data. We obtain $\logfR < -5.11 \ (95\,\% \mathrm{\ credible\ level})$, which is an order of magnitude better than the results from Ref.~\cite{Artis24} and $62\,\%$ weaker in $f_{R0}$ than our baseline result.
These different results show that a reliable $f(R)$ HMF model is needed to obtain accurate constraints on this modified gravity model. Note that different works apply different priors on \logfR\, which affects this parameter's upper limit due to the infinite volume below the lower prior. Therefore, a comparison is always affected by the prior choice when using a uniform prior in \logfR. 

The strongest constraints from large-scale probes of $f(R)$ gravity are derived from weak-lensing peak abundance using weak-lensing data from CFHTLenS \cite{Xiangkun16}. In this study, the authors found an upper bound of $\logfR < -5.16 \ (95\,\% \mathrm{\ credible\ level})$ with priors from the Planck\,15 analysis on $\Omega_{\rm m}$ and $A_s$. This result is comparable, but slightly weaker than the constraints presented here. 

\begin{figure}
        \centering
        \includegraphics[width=\linewidth]{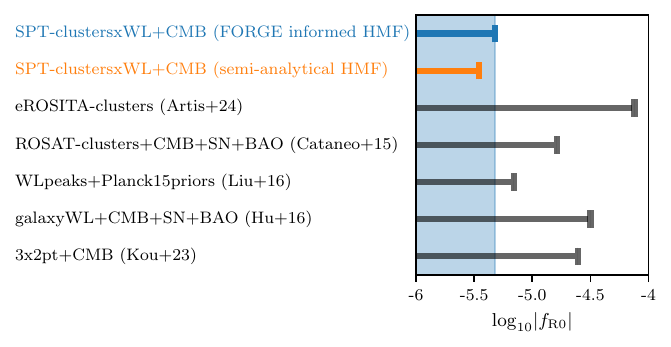}
        \vskip-0.25cm
        \caption{Comparison of \logfR\ constraints from SPTclusters + CMB with the 
         FORGE calibrated HMF (top bar) and the semi-analytical HMF model (second bar) with other recent results on cosmological scales. All limits are given at the 95\,\% credible level.}
        \label{fig:fR0_comparison}
    \end{figure}

Other recent cosmological constraints on $f(R)$ gravity have been derived from the cross-correlation of galaxies from BOSS combined with primary CMB and lensing data, yielding $\logfR \leq -4.61$ \cite{Kou23} and a combination of galaxy weak-lensing shear from the Canada–France–Hawaii Telescope Lensing Survey, CMB, SN and BAO was used to obtain $\logfR \leq -4.50$ \cite{Xiangkun16}. Figure~\ref{fig:fR0_comparison} shows the comparison of the constraints from the different cosmological probes discussed in this section. 

Modified gravity models such as $f(R)$ gravity generally enhance structure formation and thus lead to larger values of $\sigma_8$. Therefore, they do not provide a solution to the $S_8$ tension. Additionally, galaxy clusters are not sensitive to $H_0$, so our analysis does not offer any insights into the $H_0$ tension.

\section{\label{sec:summary}Summary}
This work presents constraints on $f(R)$ gravity derived from the DES and HST weak lensing informed SPT cluster abundance combined with primary CMB data from Planck\,2018. We use a sample of $1,005$ galaxy clusters selected from the SPT-SZ, SPTpolECS, and SPT500d \citep{Bleem15,Bleem20,Klein24,Bleem24} surveys with redshifts $z > 0.25$. 688 of these clusters have weak-lensing information from DES \citepalias{Bocquet24Ia,Bocquet24II} and 39 from HST \cite{Schrabback18,Schrabback21,Sommer22}. Our analysis framework is based on the methodology established by \citetalias{Bocquet24Ia} and in the recent $f(R)$ gravity forecast for upcoming Stage-III and -IV surveys \citepalias{Vogt24}.

$f(R)$ gravity alters gravity and leads, compared to GR, to a scale-dependent enhancement of structure formation and thus modifies the HMF. This enhancement in the abundance of massive galaxy clusters makes cluster samples, such as those from the SPT surveys, powerful probes for testing modified gravitational models like $f(R)$ gravity.

To capture the effects of $f(R)$ gravity on the HMF, we employ a semi-analytical approach for calculating the mass-dependent spherical collapse threshold $\delta_{\rm crit}$ \cite{Lombriser13}. The $f(R)$ HMF is then given by the GR HMF scaled by an enhancement factor, which includes the mass-dependent spherical collapse threshold. We emphasize that this model is designed to also capture the nontrivial screening effects in this modified gravity model, which play a major role at small modified gravity amplitudes that our cluster sample is able to probe.
We compare the predictions of the HMF from this semi-analytical approach with those from the FORGE simulations \cite{Arnold21}, and we find a discrepancy between the two that depends on cosmology, redshift and cluster mass. We use the simulations to calibrate the semi-analytical model to obtain a more robust HMF, while still allowing for an analysis within a broader parameter range than the simulations allow.

In this analysis, we neglect the $f(R)$ gravity effects on the gravitational lensing potential, which are subdominant compared to the weak-lensing mass calibration uncertainties derived from GR simulations \citep{Grandis21}. Furthermore, modifications to the observable--mass relations and the halo profiles are minimal, keeping the weak-lensing parameters similar to those in GR within their uncertainties \cite{Schmidt2010,Mitchell18,Ruan24}.

We achieve consistent results in all parameters compared to the $\Lambda$CDM analysis presented in \citetalias{Bocquet24II}. This is reassuring and expected, since our constraints on \logfR\ do not indicate a preference for a deviation from the $\Lambda$CDM cosmology.

We report an upper bound of $\logfR < -5.32$ at the 95\% credible level with the HMF calibrated by the FORGE simulations. A slightly tighter constraint of $\logfR < -5.46$ is obtained when using the semi-analytical HMF. 
The difference in the constraints shows the necessity for a reliable $f(R)$ HMF to place accurate constraints on $f(R)$ gravity.
The constraints reported here are the tightest constraints on $f(R)$ gravity from clusters and on cosmological scales published to date.

Upcoming Stage-III and Stage-IV surveys, such as SPT-3G \cite{Benson14} and CMB-S4 \cite{Abazajian19} or the Simons Observatory \cite{Ade19}, will provide significantly larger cluster samples, which cover a broader redshift range \cite{Raghunathan22}. In combination with next-gravitational lensing data like from the Euclid \cite{Laureijs11,Scaramella22} satellite or the Vera C.\ Rubin observatory \cite{Ivezic08,Mandelbaum18}, these cluster data sets will lead to improved constraints on $f(R)$ gravity \citepalias{Vogt24}

\begin{acknowledgments}
This research was supported by 1) the Excellence Cluster ORIGINS, which is funded by the Deutsche Forschungsgemeinschaft (DFG, German Research Foundation) under Germany's Excellence Strategy - EXC-2094-390783311, 
by 2) the Max Planck Society Faculty Fellowship program at MPE, and by 3) the Ludwig-Maximilians-Universit\"at in Munich.

The FORGE simulations and analyses of this project made use of the DiRAC@Durham facility managed by the Institute for Computational Cosmology (ICC) on behalf of the STFC DiRAC HPC Facility (\href{www.dirac.ac.uk}{www.dirac.ac.uk}). The equipment was funded by BEIS capital funding via STFC capital grants ST/K00042X/1, ST/P002293/1, ST/R002371/1 and ST/S002502/1, Durham University and STFC operations grant ST/R000832/1. DiRAC is part of the National e-Infrastructure.

BL is supported by STFC via Consolidated Grant ST/X001075/1

The Innsbruck authors acknowledge support provided by the Austrian Research
Promotion Agency (FFG) and the Federal Ministry of the Republic of
Austria for Climate Action, Environment, Mobility, Innovation and
Technology (BMK) via the Austrian Space Applications Programme
with grant numbers 899537, 900565, and 911971.

The South Pole Telescope program is supported by the National Science Foundation (NSF) through the Grant No. OPP-1852617 and 2332483. Partial support is also provided by the Kavli Institute of Cosmological Physics at the University of Chicago.
Work at Argonne National Lab is supported by UChicago Argonne LLC, Operator of Argonne National Laboratory (Argonne). Argonne, a U.S. Department of Energy Office of Science Laboratory, is operated under contract no. DE-AC02-06CH11357

Funding for the DES Projects has been provided by the U.S. Department of Energy, the U.S. National Science Foundation, the Ministry of Science and Education of Spain, 
the Science and Technology Facilities Council of the United Kingdom, the Higher Education Funding Council for England, the National Center for Supercomputing 
Applications at the University of Illinois at Urbana-Champaign, the Kavli Institute of Cosmological Physics at the University of Chicago, 
the Center for Cosmology and Astro-Particle Physics at the Ohio State University,
the Mitchell Institute for Fundamental Physics and Astronomy at Texas A\&M University, Financiadora de Estudos e Projetos, 
Funda{\c c}{\~a}o Carlos Chagas Filho de Amparo {\`a} Pesquisa do Estado do Rio de Janeiro, Conselho Nacional de Desenvolvimento Cient{\'i}fico e Tecnol{\'o}gico and 
the Minist{\'e}rio da Ci{\^e}ncia, Tecnologia e Inova{\c c}{\~a}o, the Deutsche Forschungsgemeinschaft and the Collaborating Institutions in the Dark Energy Survey. 

The Collaborating Institutions are Argonne National Laboratory, the University of California at Santa Cruz, the University of Cambridge, Centro de Investigaciones Energ{\'e}ticas, 
Medioambientales y Tecnol{\'o}gicas-Madrid, the University of Chicago, University College London, the DES-Brazil Consortium, the University of Edinburgh, 
the Eidgen{\"o}ssische Technische Hochschule (ETH) Z{\"u}rich, 
Fermi National Accelerator Laboratory, the University of Illinois at Urbana-Champaign, the Institut de Ci{\`e}ncies de l'Espai (IEEC/CSIC), 
the Institut de F{\'i}sica d'Altes Energies, Lawrence Berkeley National Laboratory, the Ludwig-Maximilians-Universit{\"a}t M{\"u}nchen and the associated Excellence Cluster Origins, 
the University of Michigan, NSF's NOIRLab, the University of Nottingham, The Ohio State University, the University of Pennsylvania, the University of Portsmouth, 
SLAC National Accelerator Laboratory, Stanford University, the University of Sussex, Texas A\&M University, and the OzDES Membership Consortium.

Based in part on observations at Cerro Tololo Inter-American Observatory at NSF's NOIRLab (NOIRLab Prop. ID 2012B-0001; PI: J. Frieman), which is managed by the Association of Universities for Research in Astronomy (AURA) under a cooperative agreement with the National Science Foundation.

The DES data management system is supported by the National Science Foundation under Grant Numbers AST-1138766 and AST-1536171.
The DES participants from Spanish institutions are partially supported by MICINN under grants ESP2017-89838, PGC2018-094773, PGC2018-102021, SEV-2016-0588, SEV-2016-0597, and MDM-2015-0509, some of which include ERDF funds from the European Union. IFAE is partially funded by the CERCA program of the Generalitat de Catalunya.
Research leading to these results has received funding from the European Research Council under the European Union's Seventh Framework Program (FP7/2007-2013) including ERC grant agreements 240672, 291329, and 306478.
We acknowledge support from the Brazilian Instituto Nacional de Ci\^encia e Tecnologia (INCT) do e-Universo (CNPq grant 465376/2014-2).

This work is based on observations made with the NASA/ESA {\it Hubble Space Telescope}, using imaging data from the SPT follow-up GO programs 12246 (PI: C.~Stubbs), 12477 (PI: F.~W.~High), 13412 (PI: T.~Schrabback), 14252 (PI: V.~Strazzullo), 14352 (PI: J.~Hlavacek-Larrondo), and 14677 (PI: T.~Schrabback).
STScI is operated by the Association of Universities for Research in Astronomy, Inc. under NASA contract NAS 5-26555.
It is also based on observations made with ESO Telescopes at the La Silla Paranal Observatory under programs 086.A-0741 (PI: Bazin), 088.A-0796 (PI: Bazin), 088.A-0889 (PI: Mohr), 089.A-0824 (PI: Mohr), 0100.A-0204 (PI: Schrabback), 0100.A-0217 (PI: Hern\'andez-Mart\'in), 0101.A-0694 (PI: Zohren), and 0102.A-0189 (PI: Zohren).
It is also based on observations obtained at the Gemini Observatory, which is operated by the Association of Universities for Research in Astronomy, Inc., under a cooperative agreement with the NSF on behalf of the Gemini partnership: the National Science Foundation (United States), National Research Council (Canada), CONICYT (Chile), Ministerio de Ciencia, Tecnolog\'{i}a e Innovaci\'{o}n Productiva (Argentina), Minist\'{e}rio da Ci\^{e}ncia, Tecnologia e Inova\c{c}\~{a}o (Brazil), and Korea Astronomy and Space Science Institute (Republic of Korea), under programs 2014B-0338 and	2016B-0176 (PI: B.~Benson).

This manuscript has been authored by Fermi Research Alliance, LLC under Contract No. DE-AC02-07CH11359 with the U.S. Department of Energy, Office of Science, Office of High Energy Physics.

\end{acknowledgments}

\appendix

\section{\label{app:comp_results} Full triangle plot}
\begin{figure*}
        \centering
        \includegraphics[width=\textwidth]{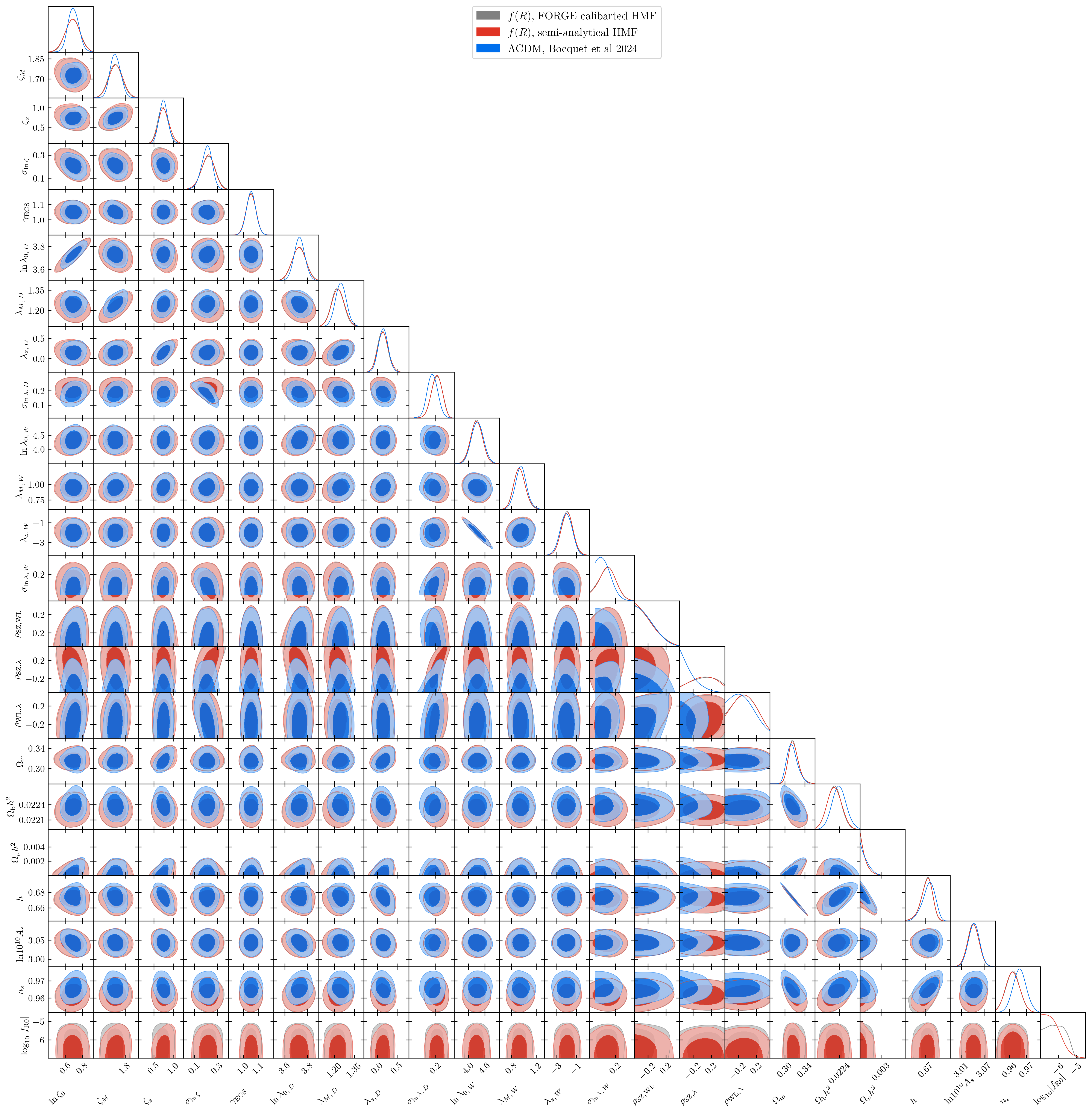}
        \vskip-0.25cm
        \caption{Posterior distribution for all parameters in the $f(R)$ and $\Lambda$CDM analyses.}
        \label{fig:comp_all}
    \end{figure*}

Figure~\ref{fig:comp_all} shows the posterior distribution for all parameters varied in the analysis, both for the $f(R)$ and $\Lambda$CDM cases of the same data \citetalias{Bocquet24II}.

\bibliography{apssamp}

\end{document}